\documentclass[12pt,a4paper,english,superscriptaddress,aps,nofootinbib]{revtex4}
\usepackage[utf8]{inputenc}
\usepackage[T1]{fontenc}
\usepackage{amsmath,amssymb,graphicx}
\makeatletter
\usepackage{babel}
\usepackage[active]{srcltx}
\usepackage{graphicx,color}
\usepackage{changebar}
\usepackage{hyperref}
\hypersetup{
    colorlinks=true,
    urlcolor= blue,
    citecolor=blue,
    linkcolor= blue}

\usepackage{amsmath}
\usepackage{braket}
\usepackage{dsfont}
\usepackage{mathtools}
\usepackage{slashed}
\usepackage{empheq}
\usepackage{tikz}
\usepackage{multirow}
\usepackage{subfigure}
\usetikzlibrary{decorations.pathmorphing}

\usepackage{graphicx}
\usepackage{amsfonts}
\usepackage{amsmath}
\begin{document}	

\title{Mitigating the information degradation in a massive Unruh-DeWitt theory}

\author{P. H. M. Barros}
\email{phmbarros@ufpi.edu.br (P. H. M. Barros)}
\affiliation{Departamento de F\'{i}sica, Universidade Federal do Piau\'{i} (UFPI), Campus Min. Petr\^{o}nio Portella - Ininga, Teresina - PI, 64049-550 - Brazil}
\author{F. C. E. Lima}
\email{cleiton.estevao@fisica.ufc.br (F. C. E. Lima)}
\affiliation{Departamento de F\'{i}sica, Universidade Federal do Cear\'{a} (UFC), Campus do Pici, Fortaleza - CE, 60455-760 - Brazil}
\author{C. A. S. Almeida}
\email{carlos@fisica.ufc.br (C. A. S. Almeida)}
\affiliation{Departamento de F\'{i}sica, Universidade Federal do Cear\'{a} (UFC), Campus do Pici, Fortaleza - CE, 60455-760 - Brazil}
\author{H. A. S. Costa}
\email{hascosta@ufpi.edu.br (H. A. S. Costa)}
\affiliation{Departamento de F\'{i}sica, Universidade Federal do Piau\'{i} (UFPI), Campus Min. Petr\^{o}nio Portella - Ininga, Teresina - PI, 64049-550 - Brazil}

\begin{abstract}
\vspace{0.5cm}
\noindent \textbf{Abstract:} We investigated the influence of the massive scalar field on the information degradation concerning the Unruh-DeWitt (UDW) detectors. In this conjecture, we adopted a system with a finite and large interaction time. To accomplish our purpose, one examines the quantum coherence of a uniformly accelerated qubit and the probability of finding the detector in the ground state. In this framework, we consider a quantum interferometric circuit to obtain the probability, visibility, and coherence. Naturally, these measurements provide us with wave-like information. Besides, one modifies the circuit to describe the path distinguishability and the particle-like information. These results are promising, as they allow us to understand the influence of the Unruh effect on the wave-particle duality. Thus, our findings announce that the increase in the scalar field mass induces a decrease in information degradation. Finally, we noted that the information concerning the Unruh effect remains preserved when $m \geq \Omega$. Therefore, the detector cannot absorb particles with mass equal to or greater than its energy gap. These results indicate that the scalar field mass is a protective factor against information degradation for systems under high acceleration conditions.\vspace{0.35cm}

\end{abstract}
	
\maketitle	

\newpage

\section{Introduction}%
\label{intro}

An understanding of quantum theory and relativity has undoubtedly been one of the focuses of theoretical physics since the 20th century  \cite{Nambu,Dyson,Feynman,Schwinger}. Since then, one can note that special relativity and quantum mechanics built the foundations of the Quantum Field Theory (QFT). Thus, it has enjoyed surprising experimental success but is less satisfactory than general relativity and quantum mechanics. A known prediction of QFT is the Fulling-Davies-Unruh effect \cite{fulling1973,Davies_1975,unruh1976}, in which an accelerated observer will perceive a thermal bath, similar to black-body radiation, which differs from the observer at rest. In other words, the accelerated observer will find himself in a warmer environment, with temperature proportional to acceleration. Therefore, one can interpret the quantum state as a static state for the inertial observer presenting itself in thermodynamic equilibrium at the non-inertial reference frame \cite{Crispino2008}.

By adopting the QFT framework, one redesigns the variable number of particles as the particle number \cite{lancaster2014quantum}. Conversely, the treatment of the concept of particles in QFT in the gravitational background becomes ambiguous \cite{wald1994quantum}. Briefly, that occurs once the particle number is not well defined \cite{wald1994quantum}; thus, it becomes necessary to describe them in terms of the response of a particle detector \cite{davies1984particles}. Mindful of this, in 1976, Unruh proposed a theoretical model of a ideal detector \cite{unruh1976}. In its proposal, Unruh suggests a detector as a punctual profile interacting with quantum fields and identifying particles transiting from their ground state to an excited state. Three years later, DeWitt suggests a simplification, i.e., he proposes a two-level monopole detector, viz., Unruh-DeWitt (UDW) detector \cite{DeWitt1980}. Although the UDW detector does not solve the question concerning the particle number in curved space, it provides a strong theoretical foundation and a practical approach to bypass this ambiguity \cite{fulling1989aspects,parker2009quantum}.

Naturally, there are some methods of measuring the particles created due to the Unruh effect. For instance, by using the Bose-Einstein condensates or optical systems, it is possible to attempt to simulate the effects of extreme acceleration, which could lead to the generation of radiation analogous to Hawking radiation \cite{Unruh2}. This approach would provide an appropriate experimental environment for investigations on the Unruh effect. Alternatively, the quantum coherence approach has been gaining prominence in the jungle of methods for measuring the particles manufactured by the Unruh effect \cite{tian2012unruh,Wang2016,he2018multipartite,Nesterov2020,Zhang2022,huang2022,Harikrishnan2022,xu2023decoherence,Pedro2024}. In the quantum coherence approach, the UDW detector absorbs particles, altering its eigenstates and causing a degradation of quantum coherence \cite{tian2012unruh,Wang2016,he2018multipartite,Nesterov2020,Zhang2022,huang2022,Harikrishnan2022,xu2023decoherence,Pedro2024}.  

Recently, we found some scenarios that amplify the degradation of quantum coherence. For instance, one notes the degradation by considering the vacuum as a dispersive medium \cite{barros2024dispersive}; furthermore, one can verify this degradation in a gravitational wave background \cite{barros2024detecting}. Another way to measure this effect is through the quantum interferometric circuit \cite{Costa2020interferometry,GoodingUnruh2020interferometric,Lopes2021interference,Pedro2024,barros2024detecting}, in which the particles created have the probability amplitude, the visibility, and the interference pattern affected. Naturally, this interferometric system allows us to test the Leggett-Garg inequality \cite{Souza_2011}, simulate Fano-Anderson problems \cite{Negrevergne2005}, and discrete Wigner functions \cite{Leonhardt1995}. Also, one can adapt the interferometer circuit to study the particle-like properties of the system through which path distinguishability gives us information about the qubit path  \cite{Jaeger1995,Englert1996,miniatura2007path,bera2015duality}. Thus, one can obtain the complementarity relation by considering the wave-like (visibility) and particle-like (which-path distinguishability) information. Consequently, this approach allows us to study the influence of the Unruh effect on the wave-particle duality \cite{Pedro2024}.

We noted, in the literature, several studies on the Unruh effect in massless quantum theories. For instance, by adopting theories of massless scalar fields, studies on information have investigated the steering and Bell nonlocality harvested by adopting UDW detectors with gravitational waves omnipresent \cite{SHWuRev1}. In this framework, one employs the UDW detectors to analyze the system's quantum coherence and quantum entanglement \cite{SHWuRev2,SHWuRev3}. Furthermore, when considering the effect of acceleration on genuine multipartite entanglement, it is noted that for UDW detectors coupled to a massless scalar field, entanglement may not show a defined correspondence with Unruh and anti-Unruh effects \cite{SHWuRev4}. Motivated by recent advancements in research on these topics, we will investigate the impact of a massive scalar field on information degradation. Thus, we regard this as motivation to study the Unruh effect by adopting massive quantum theory and using the UDW detectors for the first time. Naturally, we were also motivated to conduct this study based on several applications concerning the massive quantum theories \cite{Huang2018,Prokopec2023,dickinson2024new,zhou2021entanglement}, e.g., studies on the bosons mediating electroweak interactions, dissipative processes in neutrino oscillations \cite{benatti2000open,oliveira2010quantum}, scalar fields as components of the photon field \cite{tu2004mass,goldhaber2010photon}, and photon propagation in a waveguide \cite{rivlin1993nonzero,mendoncca2000field}.

In this article, we will investigate the effect produced by the mass of the scalar field on information degradation. To accomplish our purpose, we will study the uniformly accelerated qubit, the quantum interferometric circuit, and the path information to understand the influence of the mass in the process. We outline this work in six sections. In Sect. \ref{sec:2}, one exposes the transition rates concerning the UDW detector considering the massive scalar field. In Sect. \ref{sec:3}, the quantum coherence of the accelerated qubit and the probability (concerning the ground state) is announced. Posteriorly, we study the wave-particle duality under the Unruh effect considering the massive scalar theory and a quantum interferometric circuit, see Sect. \ref{sec:4}. Meanwhile, we show our numerical results in Sect. \ref{sec:5}. Finally, in Sect. \ref{sec:6}, our findings are announced.

\section{Transition rates of an accelerated detector}
\label{sec:2}

For our study, let us start by considering a UDW detector, i.e., a point-like detector with two-level states, i.e., ground state $\vert g\rangle$ and excited state $\vert e\rangle$. In this conjecture, the detector interacts with a massive scalar field $\phi_{m}[x(\tau)]$ through a monopole-like interaction. Allow us to consider a detector travels through a world line $x(\tau)$ with field initially in the Minkowski vacuum state $|0_{\mathcal{M}}\rangle$\footnote{Here, $\phi_{m}[x(\tau)]$ represents the massive scalar field, $m$ denotes the mass, $\tau$ is the proper time, and  $\mu(\tau)$ is the operator of the detector's monopole momentum.}. Thereby, we write the Hamiltonian for this interaction as\footnote{Conveniently,  one assumes $c = \hbar = k_B = 1$, and the metric signature is $(+,-,-,-)$.}
\begin{eqnarray}
  \mathcal{H}_{\mathrm{int}} = \lambda\chi(\tau)\mu(\tau)\phi_{m}[x(\tau)],
  \label{Hint}
\end{eqnarray}
where $\lambda$ is a coupling constant, and $\chi(\tau)$ is the switching\footnote{The switching function describes how the detector interacts with the quantum field over a finite time. So, we understand this function as a temporal window during which the UDW detector is activated and coupled to the massive field. For more details, see Ref. \cite{SFullingPDavies1976}.} function that accounts for the switch-on and switch-off of the detector. Considering the finite interaction time, this function has the properties, viz., $\chi(\tau) \approx 1$ for $|\tau| \ll T$ and  $\chi(\tau) \approx 0$ for $|\tau| \gg T$. Here, $T$ is the interaction time.

Within the first-order interaction regime, the excitation and de-excitation probabilities are $P^{\pm}_m = \lambda^2\vert\langle g\vert\mu(0)\vert e\rangle\vert^2 \mathcal{F}^{\pm}_m$, respectively. Here, $\mathcal{F}^{\pm}_m$ are response functions that carry with them the information concerning the detector's response. Mathematically, the function $\mathcal{F}^{\pm}_m$ are
\begin{eqnarray} 
\mathcal{F}^{\pm}_{m} &=& \int_{-\infty}^{\infty} d\tau  \int_{-\infty}^{\infty} d\tau' \chi(\tau)\chi(\tau') e^{\pm i\Omega(\tau - \tau')} G_{m}^{+}[x(\tau), x(\tau')],
\label{Response}
\end{eqnarray}
with $\Omega$ the angular transition frequency. If $\Omega > 0$, we have the absorption of a quantum causing an excitation in the detector. Meanwhile, for $\Omega < 0$, one has the emission of a quantum causing the de-excitation in the detector. Furthermore, the $G_{m}^{+}[x(\tau),x(\tau')]$ is the Wightman function defined as
\begin{eqnarray} 
G_{m}^{+}[x(\tau), x(\tau')] = \langle 0_{\mathcal{M}}|\phi_{m}[x(\tau)]\phi_{m}[x(\tau')]|0_{\mathcal{M}}\rangle.
\label{defWightman}
\end{eqnarray}
Note that Eq. \eqref{defWightman} is invariant under time translation for inertial and accelerated trajectories, i.e., $G_{m}^{+}[x(\tau), x(\tau')] = G_{m}^{+}(\tau - \tau')$ \cite{letaw1981quantized,padmanabhan1982general}.

By recalling the property that Wightman's function satisfies [see Refs. \cite{letaw1981quantized, sriramkumar1996finite}], viz., 
\begin{align}
	\label{Rev1}
		f(u)\left[\text{e}^{-i\Omega u}\, G_{m}^{+}(u)\right]=f\left(-\frac{\partial}{\partial\Omega}\right)\left[\text{e}^{-i\Omega u}G^{+}_{m}(u)\right],
\end{align}
where $f(u)$ is an arbitrary function with the well-defined Taylor series expansion at $u=0$. Naturally, an asymptotic expression for the transition probability for any analytic window function is
\begin{align}
    \label{Rev2}
        \mathcal{F}^{\pm}_{m}=\chi\left(i\frac{\partial}{\partial \Omega}\right)\chi\left(-i\frac{\partial}{\partial \Omega}\right) \mathcal{F}^{\pm}_{m}(\infty).
\end{align}
Therefore, the asymptotic transition probability $\mathcal{F}^{\pm}_{m}(\infty)$ boils down to 
\begin{align}
    \label{Rev3}
        \mathcal{F}^{\pm}_{m}(\infty)=\int_{-\infty}^{\infty}\,d\tau\,\int_{-\infty}^{\infty}\,d\tau'\, \text{e}^{\pm i\Omega(\tau-\tau')}G^{+}(\tau-\tau').
\end{align}

Now, let us expand the function $\chi(\tau)$ into a Taylor series around $\tau=0$, assuming $\chi(0)=1$ and $\chi'(0)=0$, that leads us to 
\begin{align}
    \label{Rev4}
        \mathcal{F}^{\pm}_{m} \approx \mathcal{F}^{\pm}_{m}(\infty)-\chi''(0)\frac{\partial^2\mathcal{F}^{\pm}_{m}(\infty)}{\partial\Omega^2}.
\end{align}
Hence, one defines the transition probability per unit time as
\begin{align}
    \label{Rev5}
        \mathcal{R}^{\pm}_{m} \approx \mathcal{R}^{\pm}_{m}(\infty)-\chi''(0)\frac{\partial^2 \mathcal{R}_{m}^{\pm}(\infty)}{\partial \Omega^2}, 
\end{align}
where
\begin{align}
    \mathcal{R}^{\pm}_{m}(\infty)=\int_{-\infty}^{\infty}\, d\Delta\tau\,\text{e}^{\pm i\Omega \Delta\tau}G^{+}(\Delta \tau) \hspace{0.3cm} \text{with} \hspace{0.3cm} \Delta \tau=\tau-\tau'.
\end{align}
One highlights that the transition probability depends on the derivatives of the window function. Consequently, if switched on and off abruptly, the detector, these contributions may lead to divergent responses. Therefore, we adopt a specific window function profile to avoid such divergences while maintaining invariance under time translation for accelerated and inertial trajectories. Naturally, these requirements lead us to use a Gaussian window function in the UDW detector, offering technical advantages that help control the window width by regulating the integrals. Thus, we assume the Gaussian window function $\chi(\tau)=\text{e}^{-\frac{\tau^2}{2T^2}}$ \cite{sriramkumar1996finite}. By adopting this profile for the window function and considering an interaction over a finite time interval, the expression for the transition probability per unit time is 
\begin{eqnarray} \label{R}
 \mathcal{R}_{m}^{\pm} \approx \mathcal{R}_{m}^{\pm}(\infty) + \frac{1}{2T^2}\frac{\partial^2 \mathcal{R}_{m}^{\pm}(\infty)}{\partial\Omega^2} + \mathcal{O}(T^{-4}),
\end{eqnarray}
where $\overline{\mathcal{R}}^{-}_{m}$ and $\overline{\mathcal{R}}^{+}_{m}$ are the rates of the excitation probability and the de-excitation probability, respectively. Besides, $\overline{\mathcal{R}}^{\pm}_{m}(\infty)$ is the transition rate for an infinite time detector ($T \to \infty$).

In light of all this, let us apply these results to a UDW detector along a trajectory that describes a motion with constant and uniform acceleration $a$ in the $z$ direction. i.e.,
\begin{eqnarray}
    t = \frac{1}{a} \sinh{a\tau}, \quad z = \frac{1}{a} \cosh{a\tau}, \quad x = y = 0.
    \label{trajectory}
\end{eqnarray}

Adopting these coordinates in Eq. \eqref{R} for a massive scalar field \cite{Crispino2008,dickinson2024new,NBirrell11}, the functions $\overline{\mathcal{R}}^{\pm}_{\overline{m}}(\infty)$ boils down to 
\begin{eqnarray}
    \overline{\mathcal{R}}^{-}_{\overline{m}}(\infty) = \frac{1}{2\pi}\frac{\sqrt{1-\overline{m}^2}}{\text{e}^{2\pi/\overline{a}}-1} \Theta(1-\overline{m})
    \label{Rinfty-}
\end{eqnarray}
and
\begin{eqnarray}
    \overline{\mathcal{R}}^{+}_{\overline{m}}(\infty) = \text{e}^{2\pi/\overline{a}}\,\overline{\mathcal{R}}^{-}_{\overline{m}}(\infty), 
    \label{Rinfty+}
\end{eqnarray}
respectively. Within this structure, $\Theta(1-\overline{m})$ is the Heaviside function, and $\overline{\mathcal{R}}^{\pm}_{\overline{m}} = \mathcal{R}^{\pm}_{\overline{m}}/\Omega$, $\overline{m} = m/\Omega$, and $\overline{a} = m/\Omega$ are dimensionless functions.

By substituting the Eqs. \eqref{Rinfty-} and \eqref{Rinfty+} into \eqref{R}, one obtains 
\begin{eqnarray}
    \overline{\mathcal{R}}^{-}_{\overline{m}} &=& \frac{1}{2\pi}\frac{\sqrt{1-\overline{m}^2}}{\text{e}^{2\pi/\overline{a}}-1} \Theta(1-\overline{m}) \Bigg\{ 1 - \frac{1}{2\sigma^2}\frac{\overline{m}^2}{(1-\overline{m}^2)^2} - \frac{1}{1-\overline{m}^2} \frac{2\pi}{\overline{a}\sigma^2}\frac{\text{e}^{2\pi/\overline{a}}}{\text{e}^{2\pi/\overline{a}}-1} \times\nonumber\\
    &\times& \Bigg[ 1 - (1-\overline{m}^2)\frac{\pi}{\overline{a}}\left( \frac{\text{e}^{2\pi/\overline{a}}+1}{\text{e}^{2\pi/\overline{a}}-1}\right)\Bigg] \Bigg\}
    \label{R-}
\end{eqnarray}
and
\begin{eqnarray}
    \overline{\mathcal{R}}^{+}_{\overline{m}} &=& \frac{e^{2\pi/\overline{a}}}{2\pi}\frac{\sqrt{1-\overline{m}^2}}{e^{2\pi/\overline{a}}-1} \Theta(1-\overline{m}) \Bigg\{ 1 - \frac{1}{2\sigma^2}\frac{\overline{m}^2}{(1-\overline{m}^2)^2} - \frac{1}{1-\overline{m}^2} \frac{2\pi}{\overline{a}\sigma^2}\frac{1}{e^{2\pi/\overline{a}}-1} \times\nonumber\\
    &\times& \Bigg[ 1 - (1-\overline{m}^2)\frac{\pi}{\overline{a}}\left( \frac{e^{2\pi/\overline{a}}+1}{e^{2\pi/\overline{a}}-1}\right)\Bigg] \Bigg\},
    \label{R+}
\end{eqnarray}
with $\sigma = \Omega T$, i.e., a dimensionless parameter. It is essential to highlight that the terms $\delta(1-\overline{m})/\Theta(1-\overline{m})$ and $\delta'(1-\overline{m})/\Theta(1-\overline{m})$ are disregard. These terms are negligible once $\delta(1-\overline{m})=0$ whenever $\overline{m}\neq 1$. Besides, $\delta(1-\overline{m}) = \infty$ when $\overline{m} = 1$; however, in this case, the contribution of the terms $\sqrt{1-\overline{m}^2}$ and $\Theta(1-\overline{m})$ is zero producing probability transition rates null.

Naturally, we noted that when $\overline{m} \to 0$, the results announced in Eqs. (\ref{R-}) and (\ref{R+}) recovers the results of the massless case, see Refs. \cite{sriramkumar1996finite, Pedro2024}. Furthermore, we remarked that when $\sigma \to \infty$, one returns to the outcome that the detector observes thermal radiation. These results assume finite interaction time does not indicate thermal radiation but a slight modification of it; this happens because the detector does not have enough time to come into thermal equilibrium with the quantum field.

\section{Accelerated single-qubit}
\label{sec:3}

\subsection{Setup}

In this section, we will investigate how the Unruh effect influences the quantum coherence of an accelerated single-qubit. Specifically, we seek to understand the mass effect on the coherence. To accomplish our purpose, let us consider the setup consisting of a detector and scalar quantum field with mass $m$. In this framework, the field at the Minkowski background is, initially,  in the vacuum state $\vert 0_{\mathcal{M}}\rangle$, while the detector is in a qubit state, viz.,
\begin{eqnarray}
|\psi_{\mathrm{D}}\rangle = \alpha|g\rangle + \beta|e\rangle,
\end{eqnarray}
where $\alpha=\cos{\frac{\theta}{2}}\text{e}^{i\frac{\varphi}{2}}$, $\beta=\sin{\frac{\theta}{2}}\text{e}^{-i\frac{\varphi}{2}}$, and $\vert\alpha\vert^2 + \vert\beta\vert^2=1$.  Here, $\theta \in [0,\pi]$ and $\varphi \in [0,2\pi]$ are the polar and azimuthal angles, respectively. 

At this stage, we are ready to find the density matrix $\hat{\rho}^{\mathrm{in}}_D$ for the detector quantum state $\vert\psi_D\rangle$, which leads us to 
\begin{eqnarray}
    \hat{\rho}^{\mathrm{in}}_D &=& |\psi_D\rangle \langle\psi_D| = |\alpha|^2 |g\rangle\langle g| + \alpha\beta^*|g\rangle\langle e| + \alpha^*\beta|e\rangle\langle g| + |\beta|^2|e\rangle\langle e|.
    \label{rhoinD}
\end{eqnarray}

\begin{figure}[!ht]
    \centering
    \includegraphics[scale=0.45]{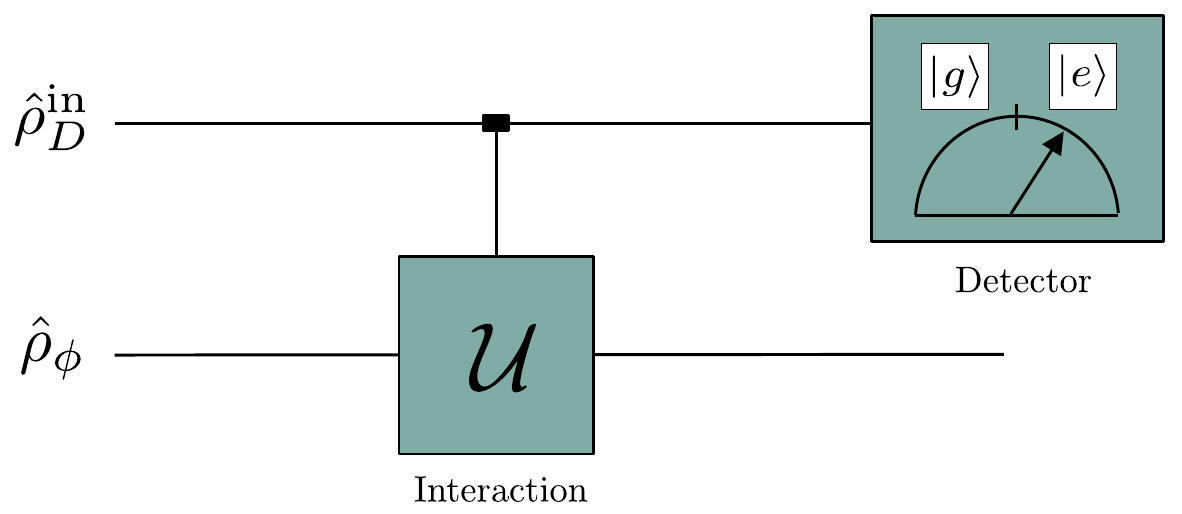}
    \caption{Schematic diagram of an accelerated qubit interacting with a massive scalar field $\phi_m$ in a finite time $T$.}
    \label{F1}
\end{figure}

We adopted a UDW detector in the qubit state. This detector will interact with a massive quantum field $\phi$ during a finite time $T$, see Fig. \ref{F1}. After this procedure, we can measure the internal states $\vert g\rangle$ and $\vert e\rangle$. These measures allow us to verify changes generated by the interaction. Mathematically, the interaction is described by $\hat{\rho}_{\mathrm{in}} = \hat{\rho}^{\mathrm{in}}_D \otimes \hat{\rho}_\phi$, where $\hat{\rho}_\phi = |0_{\mathcal{M}}\rangle \langle 0_{\mathcal{M}}|$. 

By using the Hamiltonian of the monopole interaction [Eq. \eqref{Hint}] the density matrix after the interaction boils down to
\begin{eqnarray}
    \hat{\rho}_{\mathrm{out}} &=& \mathcal{\hat{U}}^{(0)} \hat{\rho}_{\mathrm{in}} \mathcal{\hat{U}}^{(0)^\dagger} + \mathcal{\hat{U}}^{(1)} \hat{\rho}_{\mathrm{in}} + \hat{\rho}_{\mathrm{in}}\mathcal{\hat{U}}^{(0)^\dagger} + \mathcal{\hat{U}}^{(1)} \hat{\rho}_{\mathrm{in}} \mathcal{\hat{U}}^{(1)^\dagger} + \mathcal{\hat{U}}^{(2)} \hat{\rho}_{\mathrm{in}} \mathcal{\hat{U}}^{(2)^\dagger} + \mathcal{O}(\lambda^3).
\end{eqnarray}
Here, $\hat{\mathcal{U}}$ is the time evolution operator perturbated up to second order, viz.,
\begin{eqnarray}
    \mathcal{\hat{U}} = \mathcal{\hat{U}}^{(0)} + \mathcal{\hat{U}}^{(1)} + \mathcal{\hat{U}}^{(2)} + \mathcal{O}(\lambda^3),
    \label{U}
\end{eqnarray}
with
\begin{eqnarray}
    \mathcal{\hat{U}}^{(0)} &=& \mathbb{I};\\
    \mathcal{\hat{U}}^{(1)} &=& -i\lambda\int_{-\infty}^{\infty}d\tau\chi(\tau)\mu(\tau)\phi_m[x(\tau)];\\
    \mathcal{\hat{U}}^{(2)} &=& -\lambda^2\int^{+\infty}_{-\infty} d\tau \int^{+\tau}_{-\infty} d\tau' \chi(\tau)\chi(\tau')  \mu(\tau)\mu(\tau') \phi_m[x(\tau)]\phi_m[x(\tau')],
\end{eqnarray}
where $\mathbb{I}$ is the identity matrix, $\mu(\tau)=[\hat{\sigma}_{+}\text{e}^{i\Omega\tau}+\hat{\sigma}_{-}\text{e}^{-i\Omega\tau}]$. Furthermore, one defines $\hat{\sigma}_{+} = |e\rangle\langle g|$ and $\hat{\sigma}_{-} = |g\rangle\langle e|$ the creation and annihilation operators, respectively.

Now, let us examine the final configuration of the UDW detector. To perform this analysis, we must disregard the contributions due to the field configuration. That leads us to 
\begin{eqnarray}
    \hat{\rho}^{\mathrm{out}}_D = \mathrm{Tr}_{|0_{\mathcal{M}}\rangle}[\hat{\rho } _{\mathrm{out}}].
\end{eqnarray}

To obtain the density matrix, we must apply the trace to the degrees of freedom of the massive scalar field. Naturally, this procedure leads us to
\begin{equation}
    \hat{\rho}_{D,\overline{m}} = \renewcommand{\arraystretch}{1.0}\begin{pmatrix}
\hat{\rho}^{\mathrm{gg}}_{D,\overline{m}} & \hat{\rho}^{\mathrm{ge}}_{D,\overline{m}} \\
\hat{\rho}^{\mathrm{eg}}_{D,\overline{m}} & \hat{\rho}^{\mathrm{ee}}_{D,\overline{m}}
\end{pmatrix},
\label{final rho}
\end{equation}
where the components of Eq. \eqref{final rho} are
\begin{eqnarray}
\hat{\rho}^{\mathrm{gg}}_{D,\overline{m}} &=& \cos^2{\frac{\theta}{2}} + \lambda^2\left[ \sin^2{\frac{\theta}{2}} \mathcal{F}^-_{\overline{m}} - 2\cos^2{\frac{\theta}{2}} \mathbf{Re}\left(\mathcal{G}^-_{\overline{m}}\right) \right];    \\
\hat{\rho}^{\mathrm{ee}}_{D,\overline{m}} &=& \sin^2{\frac{\theta}{2}} + \lambda^2\left[ \cos^2{\frac{\theta}{2}} \mathcal{F}^+_{\overline{m}} - 2\sin^2{\frac{\theta}{2}} \mathbf{Re}\left(\mathcal{G}^+_{\overline{m}}\right) \right];    \\
\hat{\rho}^{\mathrm{eg}}_{D,\overline{m}} &=& \sin{\theta}\frac{\text{e}^{-i\varphi}}{2} + \lambda^2\left\{ \frac{\sin{\theta}}{2} \left[\text{e}^{i\varphi}\mathcal{C}^+_{\overline{m}} - \text{e}^{-i\varphi} \left(\mathcal{G}^+_{\overline{m}} + \mathcal{G}^{-*}_{\overline{m}}\right) \right] \right\}; \label{Rev_n}\\
\hat{\rho}^{\mathrm{ge}}_{D,\overline{m}} &=& \sin{\theta}\frac{\text{e}^{i\varphi}}{2} + \lambda^2\left\{ \frac{\sin{\theta}}{2} \left[\text{e}^{-i\varphi}\mathcal{C}^-_{\overline{m}} - \text{e}^{i\varphi} \left(\mathcal{G}^-_{\overline{m}} + \mathcal{G}^{+*}_{\overline{m}}\right) \right] \right\}.
    \label{rhoge}
\end{eqnarray}

In this conjecture, $\mathcal{C}^{\pm}_{\overline{m}}$, $\mathcal{G}^{\pm}_{\overline{m}}$, and $\mathcal{F}^{\pm}_{\overline{m}}$ are, respectively, 
\begin{eqnarray}
    \mathcal{C}^{\pm}_{\overline{m}} &=& \int^{+\infty}_{-\infty} d\tau \int^{+\infty}_{-\infty} d\tau' \chi(\tau)\chi(\tau') \text{e}^{\pm i\Omega(\tau+\tau')} G_{\overline{m}}(\tau, \tau'),\label{C+-}
    \\
    \mathcal{G}^{\pm}_{\overline{m}} &=& \int^{+\infty}_{-\infty} d\tau \int^{+\tau}_{-\infty} d\tau' \chi(\tau)\chi(\tau')  \text{e}^{\pm i\Omega(\tau-\tau')} G_{\overline{m}}(\tau, \tau'),\label{intG}
\end{eqnarray}
and
\begin{eqnarray}
    \mathcal{F}^{\pm}_{\overline{m}} &=& \int_{-\infty}^{\infty} d\tau \int_{-\infty}^{\infty} d\tau' \chi(\tau)\chi(\tau') \text{e}^{\pm i\Omega(\tau - \tau')} G_{\overline{m}}(\tau, \tau').\label{Flw}
\end{eqnarray}
One highlights that the terms proportional to $\lambda^2$ to have trace null. Therefore, one obtains the relation\footnote{For more details, see Refs. \cite{Martinez2014zeromode,Martinez2014,Martinez2015Causality,barros2024dispersive,barros2024detecting}.} 
\begin{eqnarray}
    \mathbf{Re}(\mathcal{G}^-_{\overline{m}}) = \frac{1}{2} \left( \mathcal{F}^-_{\overline{m}}\sin^2{\frac{\theta}{2}} + \mathcal{F}^+_{\overline{m}}\cos^2{\frac{\theta}{2}}\right),
    \label{RelationTraceless}
\end{eqnarray}
with $\mathcal{F}^{\pm}_{\overline{m}} = \sigma \overline{\mathcal{R}}^{\pm}_{\overline{m}}$. It is important to note that the reduced density matrix (\ref{final rho}) has contributions proportional to $\overline{m}$. Hence, we can verify the effects of mass in the setup of the accelerated single-qubit.

\subsection{Quantum coherence}

Quantum coherence is the property in which the superposition is omnipresent, allowing interference between different eigenstates \cite{Leggett1980}. Furthermore, it is distinguished by the preservation of the relative phases between the components of a superposition, enabling distinctive phenomena, e.g., quantum interference and entanglement \cite{streltsov2015measuring}. In this context, quantum optical methods provide a crucial set of tools for manipulating coherence \cite{Glauber1963coherent,sudarshan1963}. In our case, a two-level system, the quantum coherence between the two states $\vert g\rangle$ and $|e\rangle$ is announced by using the $l^1$ norm which by the sum of the modulus of the off-diagonal elements \cite{Baumgratz2014QuantifyingCoherence} is
\begin{align}\label{Cl1}
    \mathcal{Q}^{l^1}_{\overline{m}}(\hat{\rho}_{D,\overline{m}}) = \sum_{i \neq j} \mid \hat{\rho}_{D,\overline{m}}^{ij}\mid.
\end{align}

For our system, we exposed the off-diagonal elements in Eqs.  \eqref{Rev_n} and \eqref{rhoge}. Thus, considering the Eqs. [\eqref{Rev_n} and \eqref{rhoge}], assuming $\sigma \gg 1$ and using Eq. (\ref{RelationTraceless}), one obtains
\begin{eqnarray}
    \hat{\rho}^{\mathrm{eg}}_{D,\overline{m}} &=& \frac{1}{2}\sin{\theta} \Bigg[ \text{e}^{-i\varphi} - \sigma\lambda^2 \text{e}^{-i\varphi} \left( \overline{\mathcal{R}}^{-}_{\overline{m}}\sin^2{\frac{\theta}{2}} + \overline{\mathcal{R}}^{+}_{\overline{m}}\cos^2{\frac{\theta}{2}} \right) \Bigg],
    \label{rev_n1}
\end{eqnarray}
and 
\begin{eqnarray}\label{rev_n2}
    \hat{\rho}^{\mathrm{ge}}_{D,\overline{m}} &=& \frac{1}{2}\sin{\theta} \Bigg[ \text{e}^{+i\varphi} - \sigma\lambda^2 \text{e}^{+i\varphi} \left( \overline{\mathcal{R}}^{-}_{\overline{m}}\sin^2{\frac{\theta}{2}} + \overline{\mathcal{R}}^{+}_{\overline{m}}\cos^2{\frac{\theta}{2}} \right) \Bigg].
\end{eqnarray}

By substitution of Eqs. \eqref{rev_n1} and \eqref{rev_n2} into \eqref{Cl1}, we concluded that the $l^1$ norm coherence is 
\begin{eqnarray}
    \mathcal{Q}_{\overline{m}}^{l^1} &\approx&  |\sin{\theta}| \Bigg[ 1 - \frac{\sqrt{1-\overline{m}^2}}{2\pi} \frac{\Theta(1-\overline{m})\sigma\lambda^2}{\text{e}^{2\pi/\overline{a}}-1} \Bigg( \sin^2{\frac{\theta}{2}} + \text{e}^{2\pi/\overline{a}} \cos^2{\frac{\theta}{2}} \Bigg) \Bigg] + \mathcal{O}(\lambda^4).
    \label{coherenceQubit}
\end{eqnarray}
One highlight is that to obtain Eq. (\ref{coherenceQubit}), we adopted the condition $\sigma \gg 1$ to ensure that the detector will interact with the massive field. Also, we noted a dependence of the factor $\sqrt{1-\overline{m}}$ in Eq. \eqref{coherenceQubit} showing the mass signature in the coherence.

\subsection{On the ground state probability}

Unruh's thermal bath provides a probability amplitude degradation of the internal states in two-level systems, see Ref. \cite{Crispino2008}. Thus, we define the probabilities of these states using the reduced density matrix concerning the detector [Eq. \eqref{final rho}]. In this way, one can verify the impacts of the Unruh effect on the detector's probability amplitudes. For this analysis, one defines the probability of finding the qubit in the ground state as
\begin{eqnarray}
    P^g_{\overline{m}} = \langle g|\hat{\rho}_{D,\overline{m}}|g\rangle,
\end{eqnarray}
which leads us to
\begin{eqnarray}
    P^g_{\overline{m}} = \cos^2{\frac{\theta}{2}} + \lambda^2\left[ \mathcal{F}^-_{\overline{m}}\sin^2{\frac{\theta}{2}} - 2\mathbf{Re}\left(\mathcal{G}^-_{\overline{m}}\right)\cos^2{\frac{\theta}{2}} \right].
\end{eqnarray}

By using the constraint $\mathcal{F}^{\pm}_{\overline{m}} = \sigma\mathcal{\overline{R}}^{\pm}_{\overline{m}}$ and considering the Eq. \eqref{RelationTraceless}, one arrives at 
\begin{eqnarray}\label{Eqrrrrx}
    P^g_{\overline{m}} = \cos^2{\frac{\theta}{2}} + \sigma\lambda^2\left[ \mathcal{\overline{R}}^-_{\overline{m}}\sin^2{\frac{\theta}{2}}  - \mathcal{\overline{R}}^-_{\overline{m}}\sin^2{\frac{\theta}{2}}\cos^2{\frac{\theta}{2}} - \mathcal{\overline{R}}^+_{\overline{m}}\cos^4{\frac{\theta}{2}} \right].
\end{eqnarray}

Finally, by algebraically manipulating the Eq. \eqref{Eqrrrrx}, one obtains the probability of finding the qubit in the ground state, viz., 
\begin{eqnarray}
    P^g_{\overline{m}} = \cos^2{\frac{\theta}{2}} + \sigma\lambda^2\left[ \mathcal{\overline{R}}^-_{\overline{m}}\sin^4{\frac{\theta}{2}}  - \mathcal{\overline{R}}^+_{\overline{m}}\cos^4{\frac{\theta}{2}} \right].
    \label{ProbQubit}
\end{eqnarray}
One highlights that Eq. \eqref{ProbQubit} depends on the detector's transition rates when interacting with the massive scalar field. Therefore, the effects of the Unruh thermal bath and massive scalar field on the ground state probability become notorious. Additionally, we noted that Eq. \eqref{ProbQubit} depends on the polar angle of Bloch's sphere $\theta\,$\footnote{The Bloch sphere is a geometric representation that facilitates the visualization of quantum operations acting on the qubit. Each point on its surface corresponds to a pure state of the qubit. For further details, see Ref. \cite{Kasirajan}.}. Thus, by considering different values for $\theta$, one can examine the superposition between the internal states of the qubit.

In this conjecture, if $\theta = 0$, we have the state $|g\rangle$ concerning the north pole from the Bloch's sphere. Meanwhile, if $\theta = \pi$, we have the state $|e\rangle$ by describing the south pole from the Bloch's sphere. Naturally, intermediate points describe the state superposition. Accordingly, measurements such as quantum coherence and probability may assume an oscillatory profile once different combinations of the states $|g\rangle$ and $|e\rangle$ are allowed\footnote{That occurs once the polar angle varies in the range $\theta\in [0,\pi]$.}.

\section{Wave--particle duality}
\label{sec:4}

\subsection{Quantum scattering circuit}

\begin{figure}[!ht]
    \centering
    \includegraphics[scale=0.45]{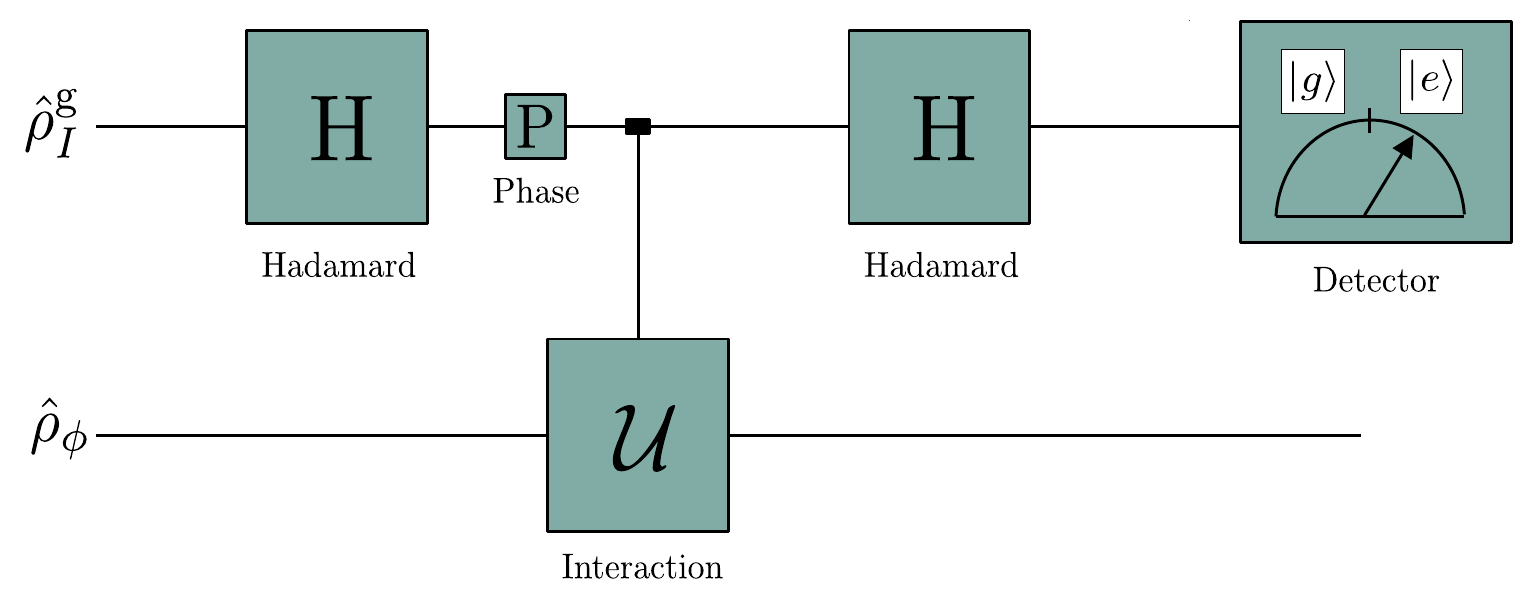}
    \caption{Schematic diagram of the quantum interferometric circuit in Unruh effect scenario \cite{barros2024detecting}.}
    \label{F2}
\end{figure}

We adopt the quantum scattering circuit (see Fig. \ref{F2}) and build a two-level setup, i.e., a single-qubit prepared according to $\hat{\rho}^{\mathrm{g}}_I$. In this conjecture, the first Hadamard gate changes the system to superposition states. Afterward, a phase accumulation occurs due to the phase-change gate $P$. By progressing, the operator $\mathcal{U}$ induces an interaction between the detector and the massive scalar field. Finally, the qubit passes through the second Hadamard gate, allowing the measurements of the detector's internal states.

The initial density matrix concerning the qubit is $\hat{\rho}^{\mathrm{g}}_I = \vert g\rangle \langle g\vert$. Therefore, after the qubit passes through the first Hadamard gate, one obtains
\begin{align}
\hat{\rho}^{\mathrm{in}}_I = \frac{1}{2}\left( \vert g\rangle \langle g\vert + \vert g\rangle \langle e\vert + \vert e\rangle \langle g\vert + \vert e\rangle \langle e\vert\right).
\end{align}
We define the initial representation as $\hat{ \rho}_{\mathrm{in}} = \hat{\rho}^{\mathrm{in}}_I \otimes \hat{\rho}_\phi$.

The probability amplitude of finding the qubit in the state $\vert g\rangle$ or $\vert e\rangle$ between the Hadamard gates has a phase accumulation $\alpha$. Therefore, the gate $P$ adds a phase if the qubit is in $\vert e\rangle$. The interaction between the detector and the massive scalar field is similar to the case of the accelerated single-qubit through the \eqref{U} operator. So, applying the second Hadamard gate, the reduced density matrix boils down to 
\begin{equation}
    \hat{\rho}_{I,\overline{m}} = \renewcommand{\arraystretch}{1.0}\begin{pmatrix}
\hat{\rho}^{\mathrm{gg}}_{I,\overline{m}} & \hat{\rho}^{\mathrm{ge}}_{I,\overline{m}} \\
\hat{\rho}^{\mathrm{eg}}_{I,\overline{m}} & \hat{\rho}^{\mathrm{ee}}_{I,\overline{m}}
\end{pmatrix},
\label{rhoI final}
\end{equation}
where
\begin{eqnarray}
    \hat{\rho}^{\mathrm{gg}}_{I,\overline{m}} &=& \cos^2{\frac{\alpha}{2}} + \frac{\lambda^2}{4} \Big[ \mathcal{F}_{\overline{m}}^- + \mathcal{F}_{\overline{m}}^+ - 8\mathbf{Re}(\mathcal{G}_{\overline{m}}^-)\cos^2{\frac{\alpha}{2}} + 2\mathcal{C}_{\overline{m}}^- \cos{\alpha} \Big]; \\
    \hat{\rho}^{\mathrm{ee}}_{I,\overline{m}} &=& \sin^2{\frac{\alpha}{2}} + \frac{\lambda^2}{4} \Big[ \mathcal{F}_{\overline{m}}^- + \mathcal{F}_{\overline{m}}^+ - 8\mathbf{Re}(\mathcal{G}_{\overline{m}}^-)\sin^2{\frac{\alpha}{2}} - 2\mathcal{C}_{\overline{m}}^- \cos{\alpha} \Big]; \\
    \hat{\rho}^{\mathrm{eg}}_{I,\overline{m}} &=& -\frac{i}{2}\sin{\alpha} + \frac{\lambda^2}{4} \Big[ \mathcal{F}_{\overline{m}}^- - \mathcal{F}_{\overline{m}}^+ + 2i\sin{\alpha \left( 2\mathbf{Re}(\mathcal{G}_{\overline{m}}^-) - \mathcal{C}_{\overline{m}}^-\right)}\Big]; \\
    \hat{\rho}^{\mathrm{ge}}_{I,\overline{m}} &=& +\frac{i}{2}\sin{\alpha} + \frac{\lambda^2}{4} \Big[ \mathcal{F}_{\overline{m}}^- - \mathcal{F}_{\overline{m}}^+ - 2i\sin{\alpha \left( 2\mathbf{Re}(\mathcal{G}_{\overline{m}}^-) - \mathcal{C}_{\overline{m}}^-\right)}\Big].
\end{eqnarray}

Naturally, by applying to  $\hat{\rho}_{I,\overline{m}}$ the traceless condition (for the contribution of $\lambda^2$), one concludes the emergence of a constraint between the coefficients $\mathcal{G}_{\overline{m}}^-$ and $\mathcal{F}_{\overline{m}}^\pm$. This constraint is 
\begin{eqnarray}
    \mathbf{Re}(\mathcal{G}_{\overline{m}}^-) = \frac{1}{4}\left( \mathcal{F}_{\overline{m}}^- + \mathcal{F}_{\overline{m}}^+\right).
    \label{tracelessrelI}
\end{eqnarray}

\subsection{Probability, visibility, and quantum coherence}

Considering the reduced density matrix of the detector [Eq. \eqref{rhoI final}], we calculate the probability of finding our system in the ground internal state $|g\rangle$. This probability is $P^{\mathrm{g}}_{I,\overline{m}} = \langle g | \hat{\rho}_{I,\overline{m}} |g\rangle$, in which leads us to 
\begin{eqnarray}
    P^{\mathrm{g}}_{I,\overline{m}} &=& \cos^2{\frac{\alpha}{2}} - \sqrt{1-\overline{m}^2} \Theta(1-\overline{m})\frac{\sigma \lambda^2 \cos{\alpha}}{8\pi} \coth{\left(\frac{\pi}{\overline{a}}\right)}.
\end{eqnarray}
Here, we use the interaction time $\sigma \gg 1$.

Naturally, we are ready to examine the interferometric visibility of the theory. Mathematically, one defines visibility as the ratio of the interference amplitude in the sum of the individual powers. Thus, the visibility is
\begin{eqnarray}
    V_{I,\overline{m}} = \frac{P^{\mathrm{g, max}}_{I,\overline{m}} - P^{\mathrm{g, min}}_{I,\overline{m}}}{P^{\mathrm{g, max}}_{I,\overline{m}} + P^{\mathrm{g, min}}_{I,\overline{m}}},
\end{eqnarray}
where the probability has the critical values when $\alpha = 0$ and $\pi$.

Explicitly, the visibility is
\begin{eqnarray}
    V_{I,\overline{m}} &\approx& 1 - \sqrt{1-\overline{m}^2} \Theta(1-\overline{m})\frac{\sigma \lambda^2}{4\pi} \coth{\left(\frac{\pi}{\overline{a}}\right)},
    \label{visibility}
\end{eqnarray}
with $\sigma \gg 1$.

Meanwhile, using the reduced density matrix \eqref{rhoI final} and considering the contributions of the off-diagonal elements, the $l^1$ norm quantum coherence is
\begin{eqnarray}
    \mathcal{Q}^{l^1}_{I,\overline{m}} &\approx& |\sin{\alpha}| \left[ 1 - \sqrt{1-\overline{m}^2} \Theta(1-\overline{m})\frac{\sigma \lambda^2}{4\pi} \coth{\left(\frac{\pi}{\overline{a}}\right)} \right].
    \label{coherenceI}
\end{eqnarray}
Therefore, we have the $l^1$ norm coherence in the setup of the quantum interferometric circuit.

\subsection{Which-path distinguishability}

\begin{figure}[!ht]
    \centering
    \includegraphics[scale=0.45]{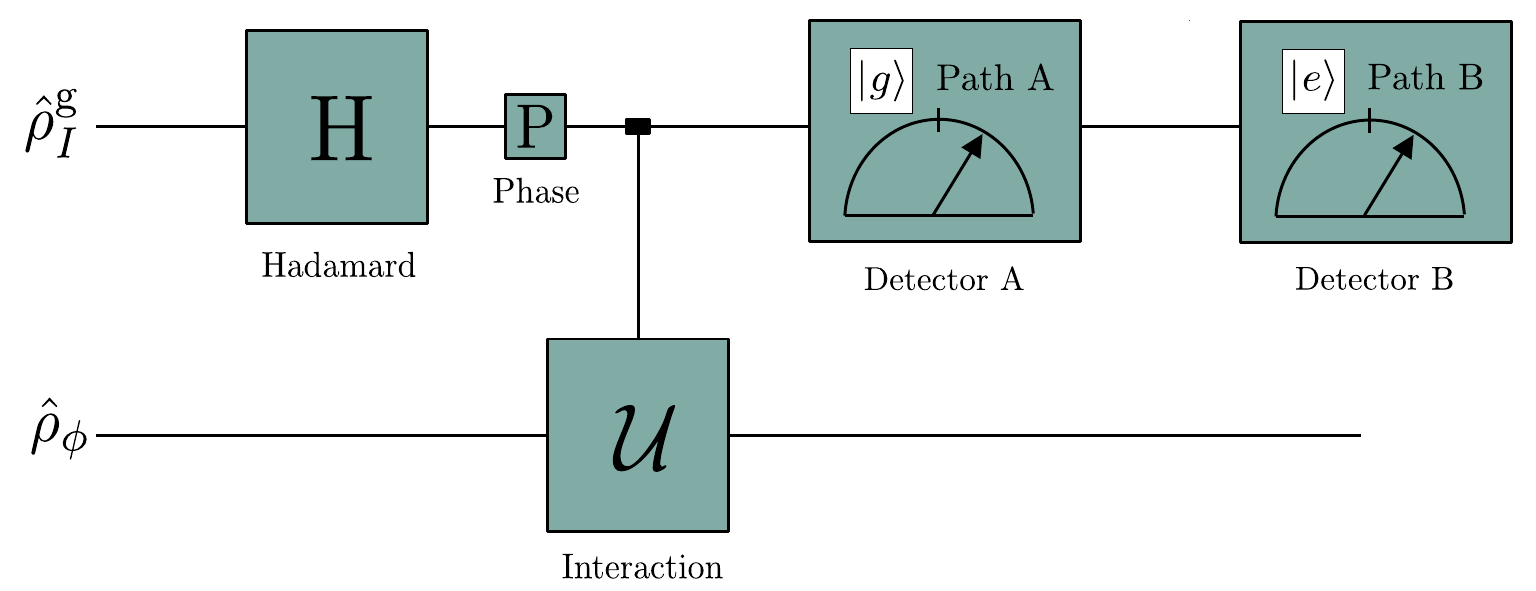}
    \caption{Schematic diagram of the path information concerning the Unruh effect.}
    \label{F3}
\end{figure}

By adopting this setup, let us search for the information path. To accomplish our purpose, we remove the second Hadamard gate from the setup (see Fig. \ref{F2}) and include two detectors, see Fig. \ref{F3}. In this scenario, one implements these two path detectors, allowing the interaction between the qubit and the massive scalar field to provide us with which path information the qubit has taken. Additionally, the internal state of the qubit is $|g\rangle$ for the detector A capturing qubits together in path A. Otherwise, these qubits reflect along path B. Thus, one can store the which-path information in the internal states of the qubit.

In this case, the reduced density matrix after the detector-field interaction corresponding to the setup (see Fig. \ref{F3}) is \begin{equation}
    \hat{\rho}_{\mathrm{Dis},\overline{m}} = \renewcommand{\arraystretch}{1.0}\begin{pmatrix}
\hat{\rho}^{\mathrm{gg}}_{\mathrm{Dis},\overline{m}} & \hat{\rho}^{\mathrm{ge}}_{\mathrm{Dis},\overline{m}} \\
\hat{\rho}^{\mathrm{eg}}_{\mathrm{Dis},\overline{m}} & \hat{\rho}^{\mathrm{ee}}_{\mathrm{Dis},\overline{m}}
\end{pmatrix},
\label{rhoDis}
\end{equation}
where
\begin{eqnarray}
    \hat{\rho}^{\mathrm{gg}}_{\mathrm{Dis},\overline{m}} &=& \frac{1}{2}  + \frac{\lambda^2}{2} \left[ \mathcal{F}^{-}_{\overline{m}} - \mathbf{Re}(\mathcal{G}^-_{\overline{m}})\right]; \\
    \hat{\rho}^{\mathrm{ee}}_{\mathrm{Dis},\overline{m}} &=& \frac{1}{2}  + \frac{\lambda^2}{2} \left[ \mathcal{F}^{+}_{\overline{m}} - \mathbf{Re}(\mathcal{G}^+_{\overline{m}})\right]; \\
    \hat{\rho}^{\mathrm{ge}}_{\mathrm{Dis},\overline{m}} &=& \frac{\text{e}^{-i\alpha}}{2}  + \frac{\lambda^2}{2} \left[ \text{e}^{i\alpha}\mathcal{C}^{-}_{\overline{m}} - \text{e}^{-i\alpha}(\mathcal{G}^-_{\overline{m}} + \mathcal{G}^{+*}_{\overline{m}})\right]; \\
    \hat{\rho}^{\mathrm{eg}}_{\mathrm{Dis},\overline{m}} &=& \frac{\text{e}^{i\alpha}}{2}  + \frac{\lambda^2}{2} \left[ \text{e}^{-i\alpha}\mathcal{C}^{+}_{\overline{m}} - \text{e}^{i\alpha}(\mathcal{G}^+_{\overline{m}} + \mathcal{G}^{-*}_{\overline{m}})\right].
\end{eqnarray} 

Again, once Eq. (\ref{rhoDis}) is traceless in terms proportional to $\lambda^2$, one attains
\begin{eqnarray}
     \mathcal{F}_{\overline{m}}^\mp - 2\mathbf{Re}(\mathcal{G}_{\overline{m}}^\pm) = 0.
    \label{tracelessDis}
\end{eqnarray}
Relevant information on this system is the path distinguishability, which characterizes the particle nature of the qubit and its quantifier through the expression
\begin{equation}
 \mathcal{D}_{\overline{m}} = \frac{|w_{A,\overline{m}} - w_{B,\overline{m}}|}{w_{A,\overline{m}} + w_{B,\overline{m}}},
 \label{Dis1}
\end{equation}
where $w_{A,\overline{m}}$ is the probability of detecting the qubit in detector A (path A) and $w_{B,\overline{m}}$ is the probability of detecting the qubit in detector B (path B).

Adopting the Eqs. (\ref{rhoDis}) and (\ref{tracelessDis}), it is written
\begin{eqnarray}
 w_{A,\overline{m}} &=& \frac{1}{2} + \frac{\lambda^2}{2} \left(\mathcal{F}_{\overline{m}}^- - \mathcal{F}_{\overline{m}}^+\right),
 \end{eqnarray}
 and
 \begin{eqnarray}
 w_{B,\overline{m}} &=& \frac{1}{2} + \frac{\lambda^2}{2} \left(\mathcal{F}_{\overline{m}}^+ - \mathcal{F}_{\overline{m}}^-\right).
\end{eqnarray}

By using $\mathcal{F}_{\overline{m}}^\pm = \sigma \overline{\mathcal{R}}_{\overline{m}}^\pm$ and Eq. \eqref{Dis1}, the path distinguishability is
\begin{equation}
 \mathcal{D}_{\overline{m}} = \sqrt{1-\overline{m}} \Theta(1-\overline{m})\frac{\sigma\lambda^2}{2\pi}.
 \label{DisFinal}
\end{equation}
This result announces the information about which path of the qubit in the setup.

\subsection{Complementarity relation}

We are now able to describe the wave-particle information for the system. The wave-particle relation quantifies the trade-off between measuring the path distinguishability of the qubit $\mathcal{D}_{\overline{m}}$ and the visibility of the interference fringe $\mathcal{V}_{\overline{m}}$ \cite{Englert1996,Englert1999,Zeilinger1999Experiment}. That definition leads us to  
\begin{equation} 
    \mathcal{C}_{I,\overline{m}} = \mathcal{V}^2_{I,\overline{m}} + \mathcal{D}^2_{\overline{m}} \leq 1.
\label{Complementarity1}
\end{equation}
Eq. \eqref{Complementarity1} is the duality relation for a Ramsey interferometer. The equal sign is valid only when the qubit and the quantum field are pure states. Meanwhile, Eq. \eqref{Complementarity1} shows a mutually exclusive relation between wave-like and particle-like behaviors. 

Finally, substituting Eqs. (\ref{visibility}) and (\ref{DisFinal}) into (\ref{Complementarity1}), we concluded
\begin{eqnarray}
    C_{I,\overline{m}} &\approx& 1 - \sqrt{1-\overline{m}^2} \Theta(1-\overline{m})\frac{\sigma \lambda^2}{2\pi} \coth{\left(\frac{\pi}{\overline{a}}\right)} + \mathcal{O}(\lambda^4),
    \label{ComplementarityFinal}
\end{eqnarray}
with $\sigma \gg 1$.

\section{Numerical results and discussion}
\label{sec:5}

\subsection{Excitation rate}

\begin{figure}[!ht]
    \centering
    \includegraphics[height=5cm,width=7.5cm]{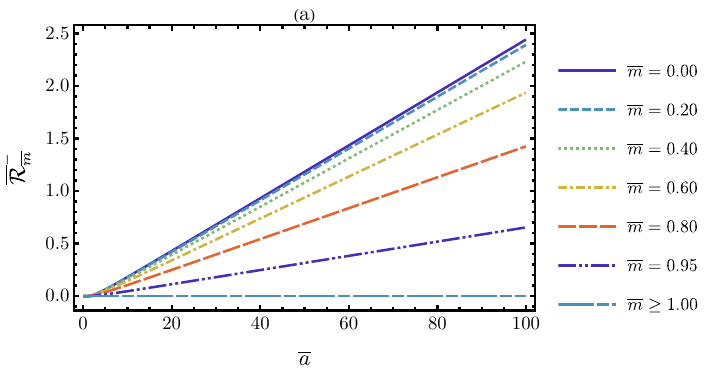}
    \includegraphics[height=5cm,width=7.5cm]{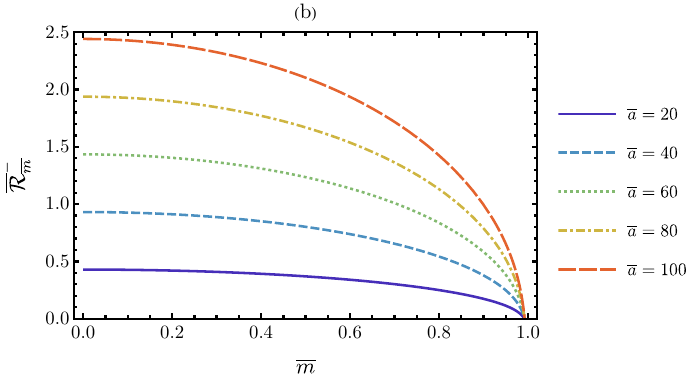}
    \caption{Excitation rate $\overline{\mathcal{R}}^{-}_{\overline{m}}$. (a) Plot of $\overline{\mathcal{R}}^{-}_{\overline{m}}$ in terms of $\overline{a}$ for several values of $\overline{m}$. (b) Plot of $\overline{\mathcal{R}}^{-}_{\overline{m}}$ in terms of $\overline{m}$ for several values of $\overline{a}$. We assume $\sigma = 10$.}
    \label{F4}
\end{figure}

In Fig. \ref{F4}(a), we exposed the excitation rate $\overline{\mathcal{R}}^{-}_{\overline{m}}$ in terms of the dimensionless parameter $\overline{a}$ for several values of $\overline{m}$. The numerical results show that the excitation rate decreases when  $\overline{m}$ increases. Additionally, in Fig. \ref{F4}(b), we plot the excitation rate $\overline{\mathcal{R}}^{-}_{\overline{m}}$ in terms of $\overline{m}$ for several values of $\overline{a}$. Thus, one notes that the transition rate decreases as $\overline{m}$ approaches one and becomes zero for $\overline{m} \geq 1$. 

The presence of the function $\Theta(1-\overline{m})$ in $\overline{\mathcal{R}}^{-}_{\overline{m}}$ implies that the absorption of a single quantum of mass $\overline{m}$ by the detector will not occur unless the energy level spacing $\Omega$ in the detector is at least equal to the rest energy of the particle. Put another way, the transition rate is zero for $\overline{m} \geq 1$, meaning the detector cannot absorb particles with mass equal to or greater than the energy gap \cite{NBirrell11,dickinson2024new}.

\subsection{Accelerated single-qubit}

\begin{figure}[!ht]
    \centering
    \includegraphics[height=5cm,width=7.5cm]{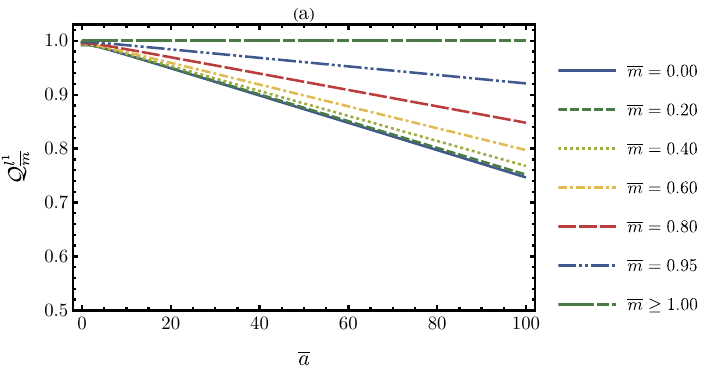}
    \includegraphics[height=5cm,width=7.5cm]{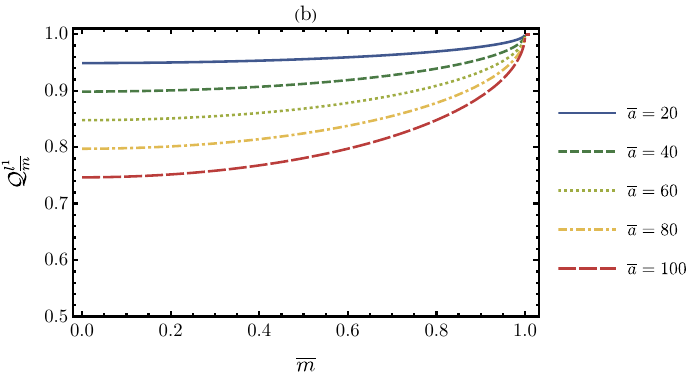}
    \includegraphics[height=5cm,width=7.5cm]{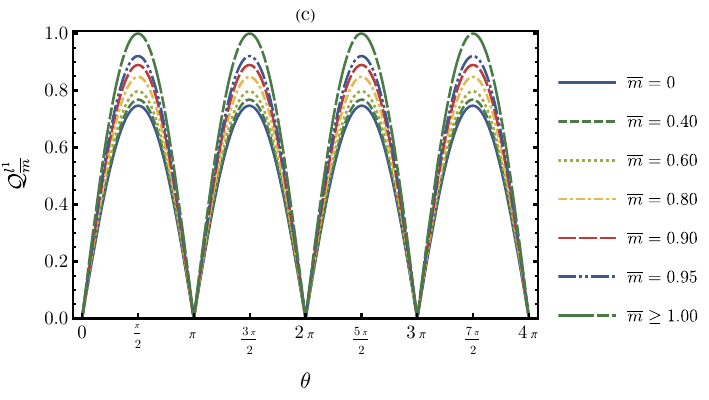}
    \caption{Numerical results of the quantum coherence $\mathcal{Q}^{l^1}_{\overline{m}}$. (a) $\mathcal{Q}^{l^1}_{\overline{m}}$ vs. $\overline{a}$ for different values of $\overline{m}$. (b) $\mathcal{Q}^{l^1}_{\overline{m}}$ vs. $\overline{m}$ for different values of $\overline{a}$. (c) $\mathcal{Q}^{l^1}_{\overline{m}}$ vs. $\theta$ for different values of $\overline{m}$. We adopted $\theta = \pi/2$, $\lambda = 0.1$, and $\sigma=10$.}
    \label{F5}
\end{figure}

In Fig. \ref{F5}, we announced the numerical results of the $l^1$ norm coherence of an accelerated single-qubit and the mass influence on it. Note that in Fig. \ref{F5}(a) and Fig. \ref{F5}(b), the coherence increases with the value of $\overline{m}$, becoming equal to one (maximum) for $\overline{m} \geq 1$, and causing mitigation of the coherence degradation. Furthermore, the excitation rate studied suggests that the detector does not absorb any particles. Thus, as the mass increases, the detector becomes restricted, absorbing gradually less when the mass increases. Consequently, when $m \geq \Omega$ (or $\overline{m} \geq 1$), the detector stops responding. In this case, the coherence degradation due to the Unruh effect will cease. Additionally, by considering Fig. \ref{F5}(c), one notes an oscillatory behavior by numerical inspection of the $l^1$ norm coherence in terms of the qubit polar angle $\theta$ for different values of $\overline{m}$.

Note that since the $l^1$ norm coherence measures the superposition capacity of the system and the qubit polar angle determines the superposition of the states. Thus, the analysis of the coherence in terms of the polar angle results in an oscillatory profile due to the mixing of the system's internal states. Consequently, the coherence amplitude is maximal when $\theta$ is a semi-integer multiple of $\pi$, which allows us to announce a maximal superposition of mixed states. Meanwhile, the oscillation amplitude is zero when $\theta$ is an integer multiple of $\pi$; here, we localize the poles from Bloch's sphere where there is no superposition. Furthermore, one notes that if $\overline{m}$ increases, the oscillation amplitude will increase, suggesting the mitigation of coherence degradation.

\begin{figure}[!ht]
    \centering
    \includegraphics[height=6cm,width=9cm]{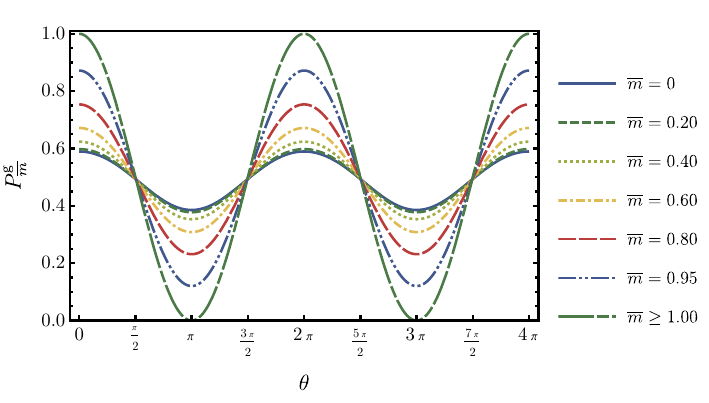}
    \caption{Probabilities $P^{\mathrm{g}}_{\overline{m}}$ vs. $\theta$ for different values of $\overline{m}$. In this case, we adopt $\lambda=0.05$, $\overline{a}=100$, and $\sigma=10$.}
    \label{F6}
\end{figure}

In Fig. \ref{F6}, we exposed the probability of finding the qubit in the ground state $P^{\mathrm{g}}_{\overline{m}}$ [Eq. \eqref{ProbQubit}] in terms of the polar angle $\theta$ for different values of $\overline{m}$. In this framework, we noted that $P^g_{\overline{m}}$ measures the probability of the ground state $\vert g\rangle$. The analysis of this probability in terms of $\theta$ yields an oscillatory profile (resembling the case of the coherence). In this case, we noted that the oscillation amplitude of the probability $P^g_{\overline{m}}$ is maximum when $\theta$ is an integer multiple of $2\pi\,$\footnote{This case describes the ground state $\vert g\rangle$, localized at the north pole from Bloch's sphere.}. On the other hand, the oscillation amplitude will be minimum when $\theta$ is an integer multiple of $\pi\,$\footnote{This case describes the excited state $\vert e\rangle$, localized at the south pole from Bloch's sphere.}. In this case, one notes increasing $\overline{m}$ an increase in oscillation amplitude. Specifically, for the case $\overline{m} \geq 1$, we have the oscillation amplitude maximum, meaning that the Unruh effect no longer influences the probability of detecting the internal states of the accelerated qubit.

\subsection{Wave-particle duality}

\begin{figure}[!ht]
    \centering
    \includegraphics[height=4.5cm,width=4.5cm]{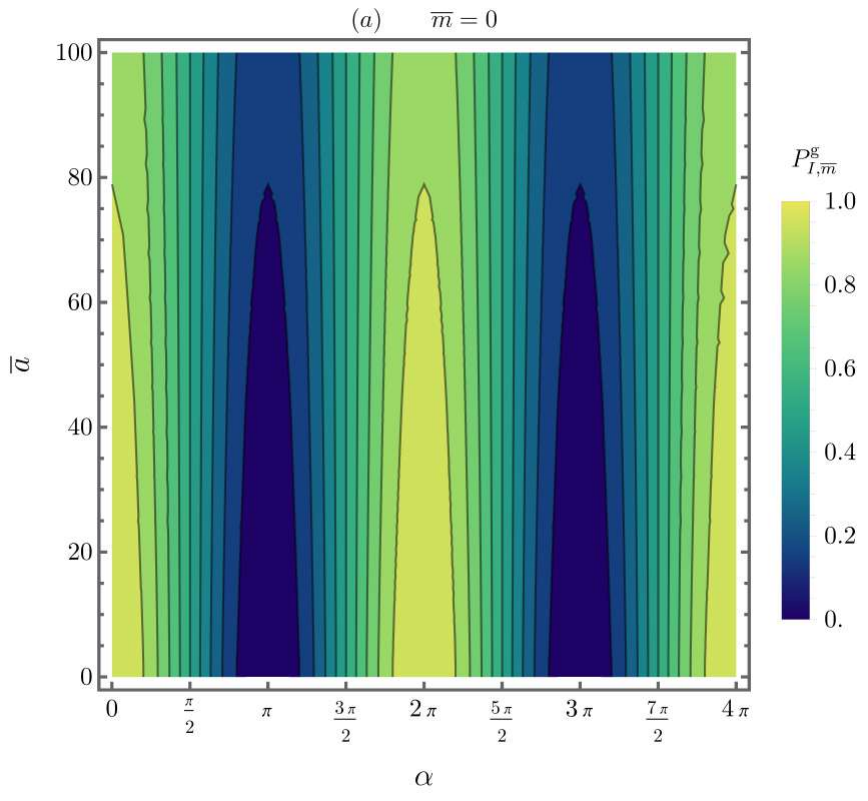}
    \includegraphics[height=4.5cm,width=4.5cm]{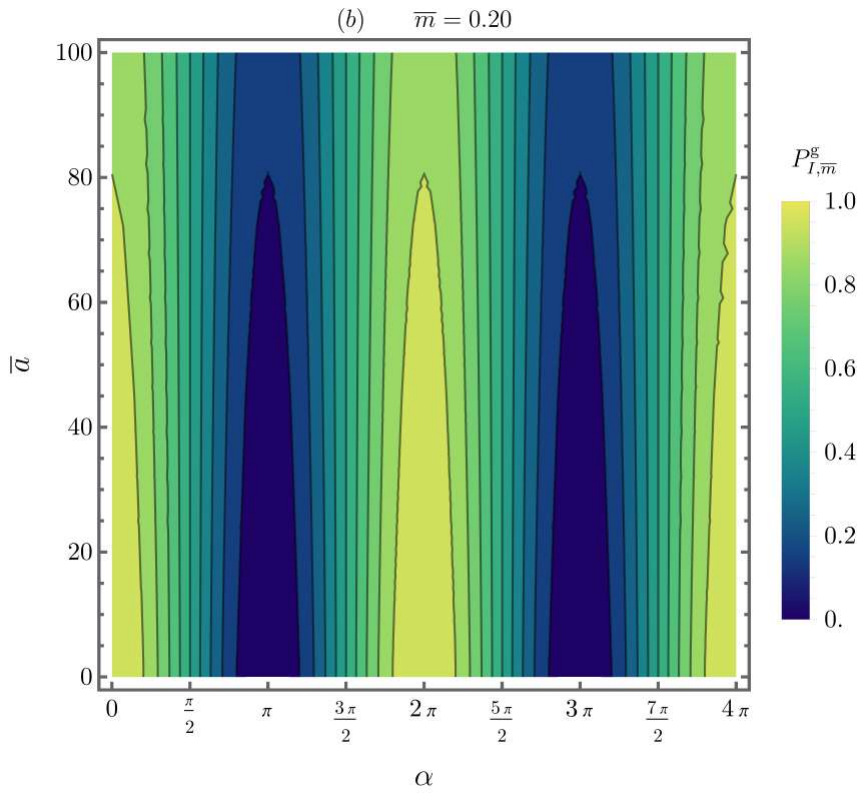}
    \includegraphics[height=4.5cm,width=4.5cm]{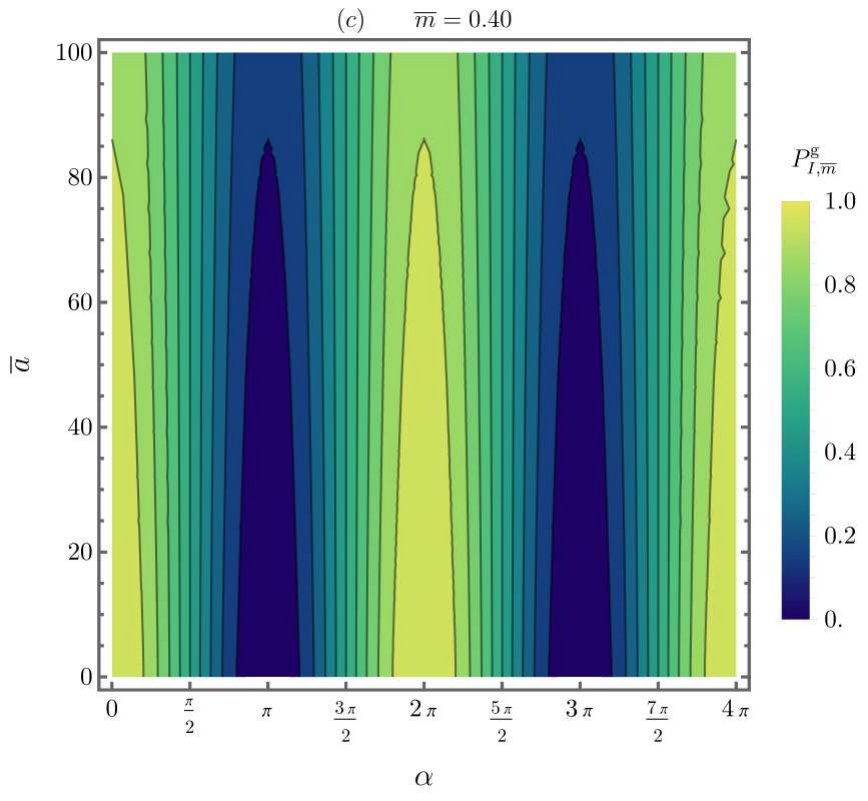}\\
    \includegraphics[height=4.5cm,width=4.5cm]{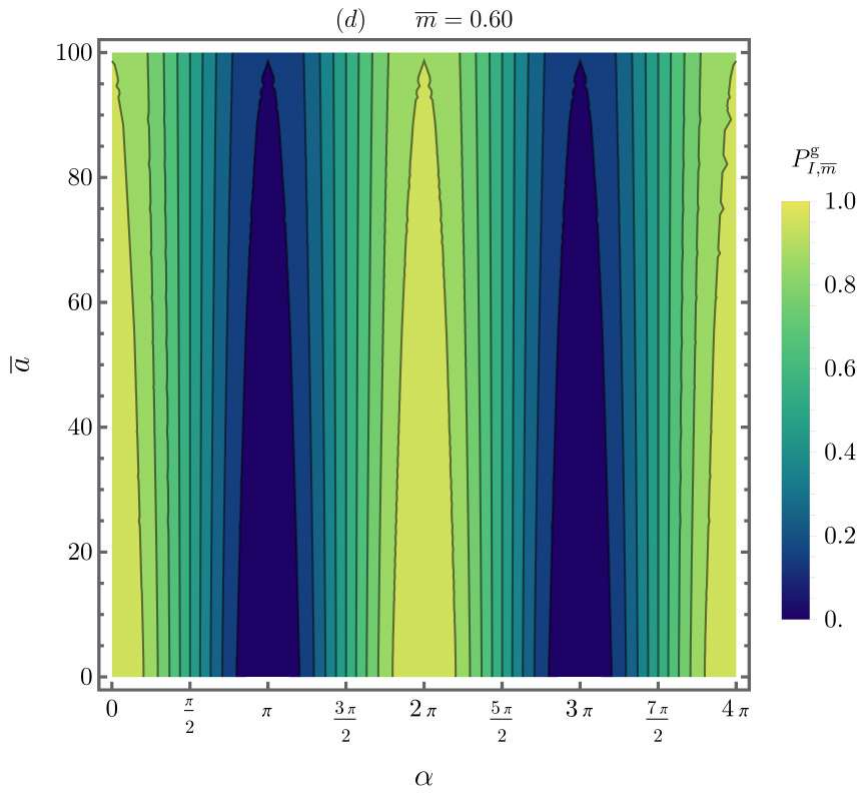}
    \includegraphics[height=4.5cm,width=4.5cm]{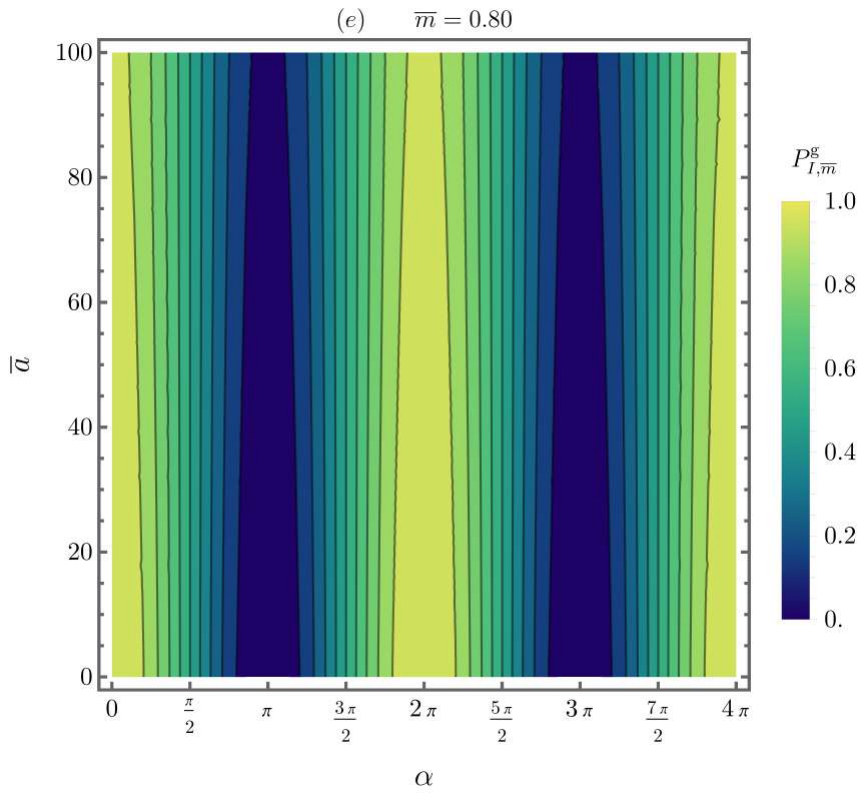}
    \includegraphics[height=4.5cm,width=4.5cm]{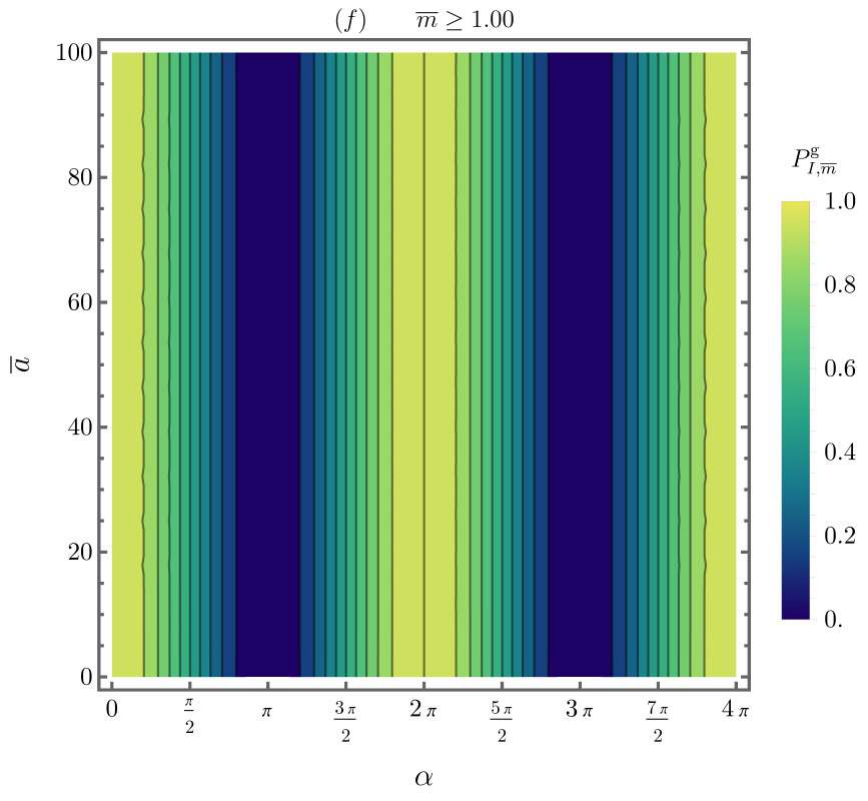}
    \caption{Probabilities $P^{\mathrm{g}}_{I,\overline{m}}$ in terms of  $\overline{a}$ and $\alpha$ adopting several $\overline{m}$. In this case, we assume $\lambda=0.1$ and $\sigma=10$.}
    \label{F7}
\end{figure}

\begin{figure}[!ht]
    \centering
    \includegraphics[height=4.5cm,width=4.5cm]{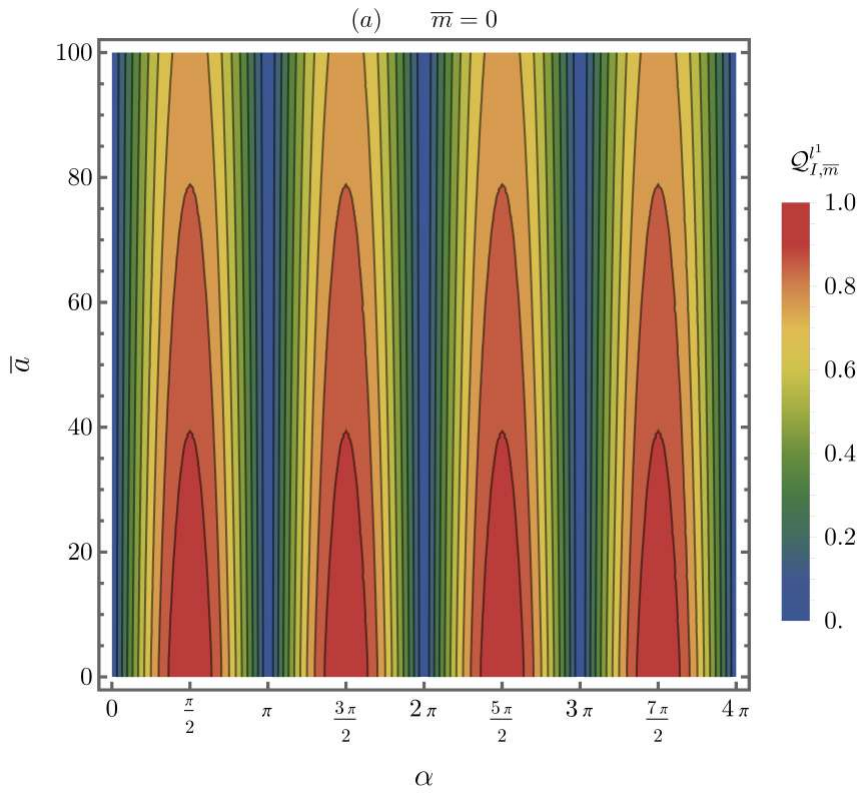}
    \includegraphics[height=4.5cm,width=4.5cm]{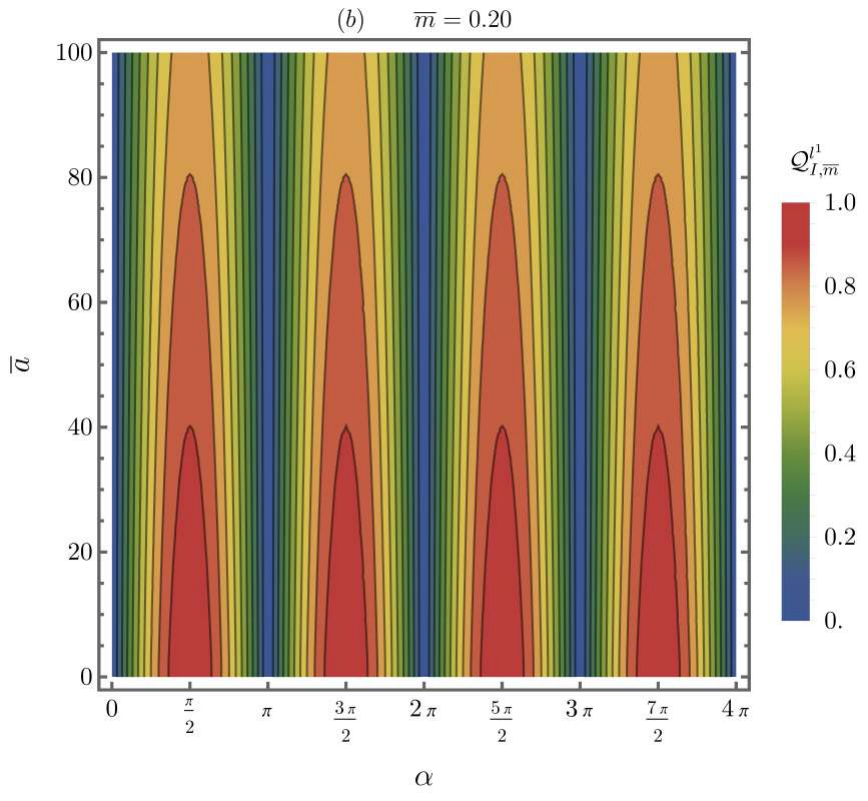}
    \includegraphics[height=4.5cm,width=4.5cm]{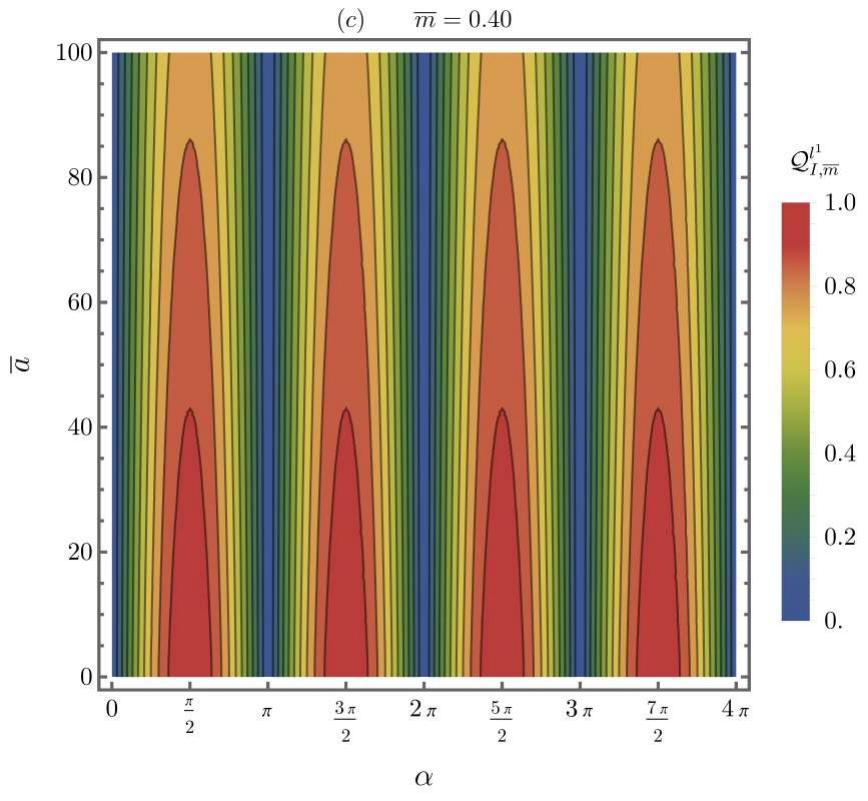}\\
    \includegraphics[height=4.5cm,width=4.5cm]{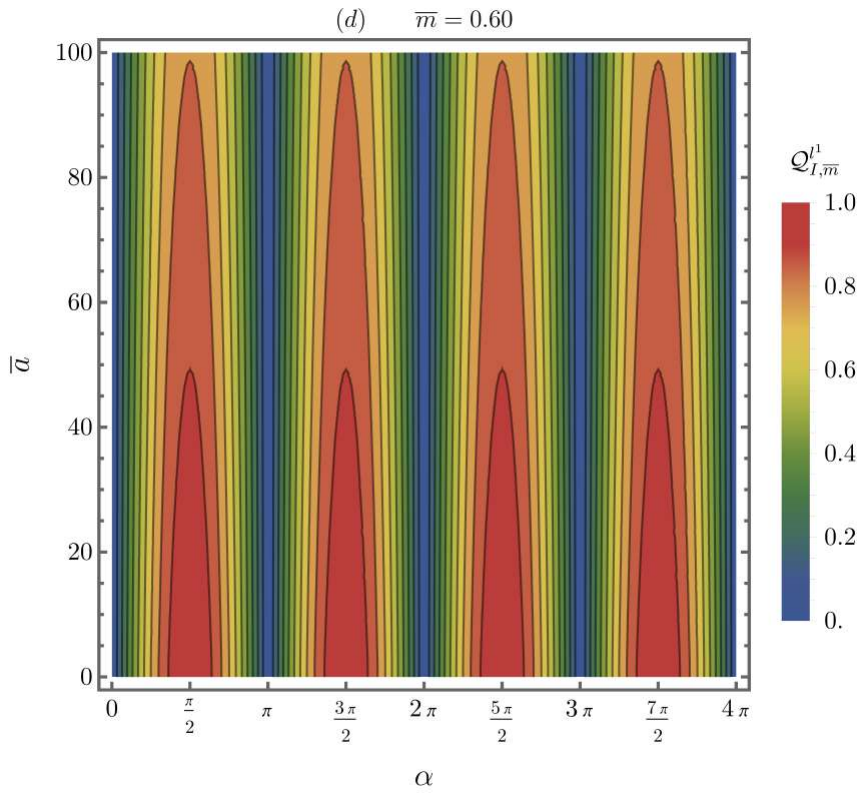}
    \includegraphics[height=4.5cm,width=4.5cm]{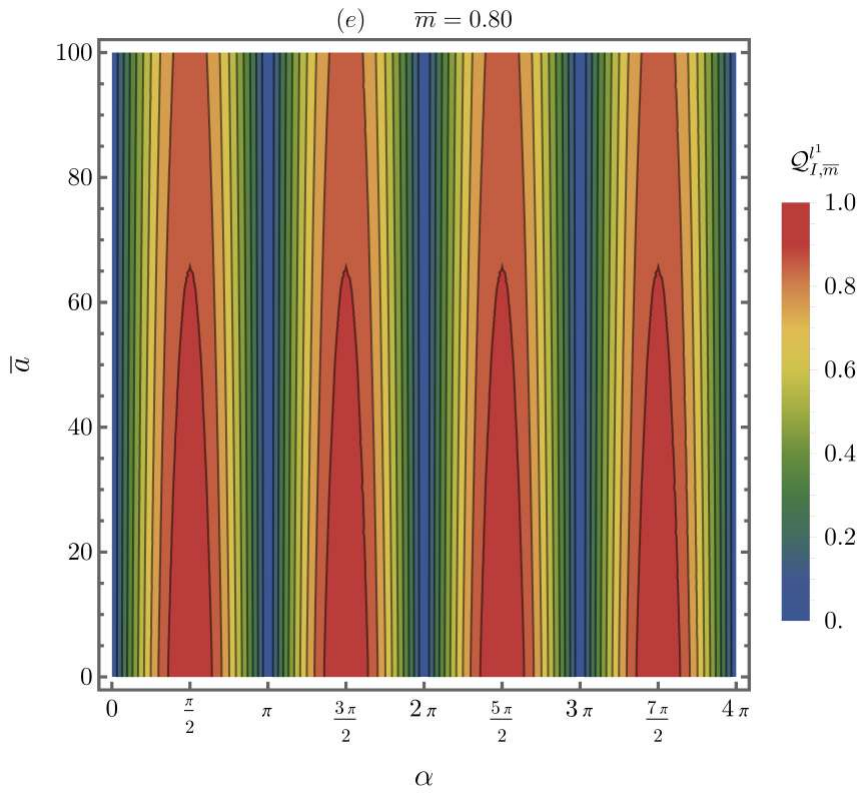}
    \includegraphics[height=4.5cm,width=4.5cm]{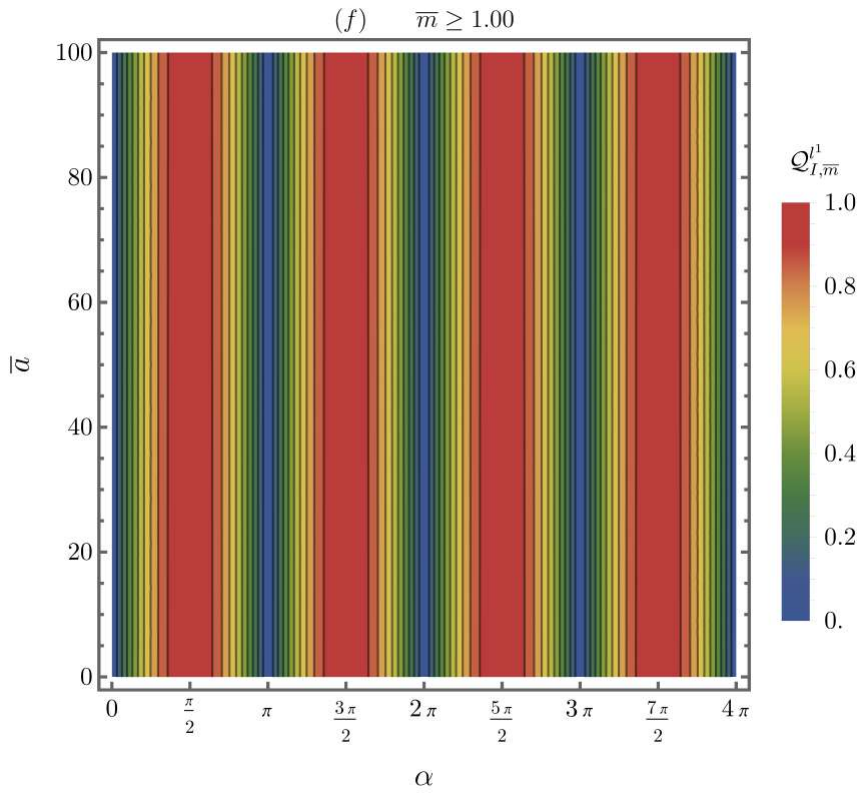}
    \caption{Quantum coherence $\mathcal{Q}^{l^1}_{I,\overline{m}}$ in terms of $\overline{a}$ and $\alpha$ assuming several $\overline{m}$.  We adopted $\lambda=0.1$ and $\sigma=10$.}
    \label{F8}
\end{figure}

The numerical results plotted in Fig. \ref{F7}(a)-\ref{F7}(e), for the phase being an integer number $\pi$ in the high acceleration regime, allow us to remark the degradation of the probability amplitude. Similarly, in Fig. \ref{F8}(a)-\ref{F8}(e), for high accelerations and semi-integer numbers of $\pi$, one concludes that the coherence degradation. Furthermore, for these results under the conditions presented, we noted that for larger values of $\overline{m}$, the probability amplitude and coherence profiles are fringes sharper of the quantum interferometric circuit. On the other hand, see that for the case $\overline{m} \geq 1$, we have a perfectly sharp image of the fringes of $P^{\mathrm{g}}_{I,\overline{m}}$ and $\mathcal{Q}^{l^1}_{I,\overline{m}}$, and therefore, this means that the detector does not respond under these conditions because it does not absorb particles with mass equal to or greater than its energy gap.

\begin{figure}[!ht]
    \centering
    \includegraphics[height=5cm,width=7.5cm]{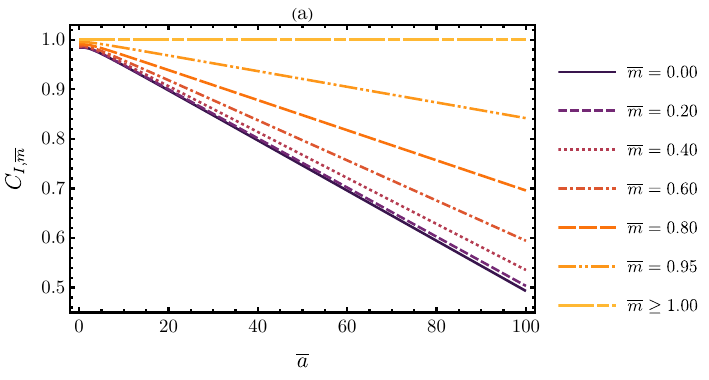}
    \includegraphics[height=5cm,width=7.5cm]{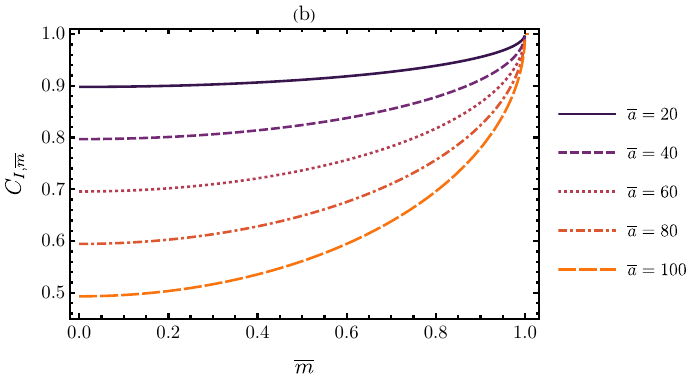}
    \caption{Complementarity relation $C_{\mathrm{m}}$. (a) $C_{\mathrm{m}}$ vs. $\overline{a}$ for different $\overline{m}$. (b) $C_{\mathrm{m}}$ vs. $\overline{m}$ for different $\overline{a}$. We adopted $\lambda=0.1$ and $\sigma=10$.}
    \label{F9}
\end{figure}

In Fig. \ref{F9}, we numerically analyzed the effects of mass on the wave-particle duality under the Unruh effect. This is done through the complementarity relation $C_{I,\overline{m}}$, which makes use of the which-path distinguishability $\mathcal{D}_{\overline{m}}$ and the interferometric visibility $\mathcal{V}_{I,\overline{m}}$. These results show that when $\overline{m}$ increases, the complementarity relation becomes greater. For the case where the mass is greater than or equal to the energy gap, one obtains that the complementarity relation is maximum ($C_{I,\overline{m}} = 1$). Besides, these findings mean that because mass causes a mitigation of quantum coherence, this implies a mitigation of the information loss of the wave-particle duality. Therefore, we concluded that mass is a factor that protects the information of the interferometric system.

\section{The limitations of the model}

Now, let us discuss briefly the limitations and assumptions omnipresent in our model. Initially, we used the perturbation theory approach to investigate the detector-field interaction. This approach assumes a small coupling constant ($\lambda \ll 1$), ensuring the validity of the perturbative expansion, but is limited to regimes of weak interactions. Furthermore, we assumed that the interaction time occurs during a finite and large time interval ($\sigma \gg 1$) compared to the transition time of the two-level system ($\sim \Omega^{-1}$), ensuring the detector-field interaction time. It is essential to highlight that, in other regimes, one can have changes that may prevent generalization to arbitrary time scales.

To accomplish our purpose, we performed a numeric investigation of the result. In this case, we assumed the high acceleration regime ($\overline{a} \gg 1$), i.e., the detector acceleration is much larger than the transition frequency $\Omega$. However, the model does not include back-reaction effects, assuming that the detector does not significantly influence the field, which may be a limitation in strongly coupled scenarios. Furthermore, higher-order corrections and non-perturbative effects are despicable.

In this conjecture, it is clear that these assumptions directly influence the generalization of the numerical results once the effects analyzed are valid only within these mentioned regimes. Therefore, aiming at more complex situations, additional, more detailed studies are necessary to evaluate more general scenarios.

\section{Summary and conclusion}\label{sec:6}

We investigated the influence of the mass of the scalar field on the information degradation in the accelerated quantum regime. Our findings indicate that the excitation rate of the detector decreases when the mass increases. Furthermore, the detector does not respond when the mass of the scalar field is equal to or greater than the detector's energy gap. The results regarding the accelerated qubit and the quantum interferometric circuit showed that the degradation of coherence and the probability amplitude decreases when $\overline{m}$ increases. These results announced a mitigation of the influences of the Unruh effect on the detector's internal states.

Through the interferometric visibility and the which-path distinguishability, we obtained the complementarity relation, which allowed us to investigate the effects of the mass on the Unruh effect on the wave-particle duality. These findings showed mitigation in the wave-like and particle-like information loss. That indicates that the mass of the scalar field plays a protective role against the thermal bath created due to high acceleration. We noted that $\overline{m} \geq 1$, the detectors do not respond. Therefore, we do not have coherence, probability amplitude, and complementarity relation degradation. These results mean that the thermal bath does not degrade the information of the high-acceleration quantum systems.

Indeed, our findings contribute to understanding the nature of relativistic quantum information. Once we announced that the existence of a massive scalar field induces mitigation of information degradation in accelerated frames. Therefore, this result suggests that information loss is not an absolute phenomenon in accelerated quantum systems but rather depends on specific characteristics of the quantum field under consideration.

Furthermore, the possibility of information properties to be protected emphasizes the necessity for further research on the Unruh effect. Thus, our findings not only advance the fundamental understanding of relativistic quantum information but also offer new directions for bridging the gap of incompatibility between quantum mechanics and general relativity. Generally speaking, these results deepen our understanding of quantum information preservation and open doors for promising technological perspectives. Naturally, this mitigation can influence the development of quantum computers, quantum communications, and high-precision quantum sensors.

\section*{ACKNOWLEDGMENT}

The authors would like to express their sincere gratitude to the Coordenação de Aperfeiçoamento de Pessoal de Nível Superior (CAPES), Conselho Nacional de Desenvolvimento Científico e Tecnológico (CNPq) and Funda\c{c}\~{a}o Cearense de Apoio ao Desenvolvimento Cient\'{i}fico e Tecnol\'{o}gico (FUNCAP) for their valuable support. P.H.M. Barros is thankful for grant No. 88887.674765/2022-00 (Doctoral Fellowship-CAPES). F. C. E. Lima and C. A. S. Almeida are supported, respectively, for grants No. 171048/2023-7 (CNPq/PDJ) and 309553/2021-0 (CNPq/PQ). C. A. S. Almeida is also supported by Project UNI-00210-00230.01.00/23 (FUNCAP). The authors thank the Referee for the constructive feedback and valuable comments.

\section*{CONFLICTS OF INTEREST/COMPETING INTEREST}

The authors declared that there is no conflict of interest in this manuscript. 

\section*{DATA AVAILABILITY}

No data was used for the research described in this article.

\end{document}